\documentclass[amsmath,nofootinbib,notitlepage,prd,twocolumn]{revtex4-2}

\usepackage[T1]{fontenc}
\usepackage[svgnames]{xcolor}
\usepackage{aas_macros,graphicx,hyperref,orcidlink,booktabs,mathrsfs,amsmath,amssymb,physics,bm,placeins,rotating,multirow,siunitx}

\hypersetup{colorlinks=true,citecolor=RoyalBlue,linkcolor=RoyalBlue,urlcolor=RoyalBlue}
\newcommand{\ed}{\mathop{}\!\mathrm{d}}

\newcommand{\ab}[1]{\left|#1\right|}

\newcommand{\br}[1]{\left[#1\right]}
\newcommand{\cu}[1]{\left\{#1\right\}}
\newcommand{\pa}[1]{\left(#1\right)}

\usepackage[normalem]{ulem}
\usepackage{mathtools}
\usepackage{enumitem}
\usepackage[caption=false]{subfig}
\usepackage{listings}
\usepackage{lmodern}
\usepackage{microtype}
\lstdefinestyle{cppstyle}{
  language=C++,
  basicstyle=\ttfamily\small,
  keywordstyle=\bfseries,
  commentstyle=\itshape,
  columns=fullflexible,
  keepspaces=true,
  showstringspaces=false,
  breaklines=true,
  frame=single
}

\newcommand{\cZk}[1]{\check{Z}^{\mathrm{\rm in}}_{#1}}
\newcommand{\bcZk}[1]{\bar{\check{Z}}^{\mathrm{\rm in}}_{#1}}

\begin{document}

\title{A Systematic Study of Resonance-Driven Flux Modifications in Extreme-Mass-Ratio Inspirals}

\author{Edoardo Levati\,\orcidlink{0009-0002-5316-3762}} 
\email{levae25@wfu.edu}
\affiliation{Department of Physics, Wake Forest University, Winston-Salem, NC 27109, USA}
\author{Lennox S. Keeble\,\orcidlink{0009-0009-5796-631X}} 
\affiliation{Department of Physics, Wake Forest University, Winston-Salem, NC 27109, USA}
\author{Alejandro C\'ardenas-Avenda\~no\,\orcidlink{0000-0001-9528-1826}} 
\affiliation{Department of Physics, Wake Forest University, Winston-Salem, NC 27109, USA}

\begin{abstract}
Transient orbital resonances can introduce phase-dependent corrections to the evolution of extreme-mass-ratio inspirals (EMRIs), potentially altering their long-term dynamics and emitted gravitational-wave signals. In this work, we quantify the resonance-induced modifications to the energy, axial angular momentum, and Carter constant fluxes and compute the corresponding resonance coefficients across a broad region of the orbital parameter space. Using the publicly available \texttt{pybhpt} code, we solve the Teukolsky equation in the frequency domain to coherently combine the degenerate radial and polar harmonics that arise at resonance. We analytically derive a selection rule governing the relative radial-polar phase dependence of the resonant flux modifications. We argue that the relative strength of the resonant flux modifications reflects a balance between symmetry-induced cancellations and the degree to which the resonant orbit samples the underlying two-dimensional orbital phase space. For the dynamically important $3{:}2$ and $2{:}1$ resonances, we also characterize how the resonance coefficients vary with the primary black-hole spin, orbital eccentricity and inclination. Our results constitute the largest set of Teukolsky-based resonance coefficients calculated to date and provide essential input for future studies of transient orbital resonances in EMRIs.
\end{abstract}

\maketitle

\section{Introduction}
\label{sec:intro}

The Laser Interferometer Space Antenna (LISA)~\cite{LISA:2024hlh} will extend gravitational-wave (GW) astronomy to mHz frequencies, bridging the gap between ground-based detectors and pulsar timing arrays. One of the primary sources for LISA will be extreme-mass-ratio inspirals (EMRIs): binary systems consisting of a stellar-mass compact object of mass $\mu$ slowly inspiraling into a supermassive black hole (BH) of mass $M$~\cite{LISA:2022yao}. Such systems, characterized by their small mass ratios $\eta\equiv \mu/M\lesssim10^{-4}$, are expected to exhibit $\sim10^{5}$ cycles in the LISA band~\cite{LISA:2024hlh}. On the one hand, this means the small body will be an excellent probe of the background spacetime of the supermassive BH, enabling precise measurements of its mass and spin and, potentially, tests of general relativity~\cite{Ryan:1995wh,Ryan:1997hg,Speri:2024qak, CardenasSopuertaReview,Barack:2018yly,Barack:2003fp,Speri:2026ade}. On the other hand, accurately modeling the complex dynamics of EMRIs over such a large number of cycles is theoretically and computationally challenging. For the purposes of parameter estimation, EMRI waveform models may require GW phase errors of less than one cycle over the time spent in the LISA band~\cite{Pound_2021,Burke:2024pfa,Amaro_Seoane_2007,FlanaganHinderer1,LISAConsortiumWaveformWorkingGroup:2023arg}.

The small mass ratios inherent to EMRI systems cause a separation between orbital ($\sim M$) and radiation-reaction ($\sim M/\eta$) timescales. This permits a two-timescale expansion of their equations of motion~\cite{Mino2005, FlanaganHinderer1, MillerPound2021}. Off resonance, the leading-order (adiabatic) secular evolution is governed by the torus-averaged first-order dissipative gravitational self-force, with sub-leading (post-adiabatic) effects from the first-order conservative self-force, the oscillatory first-order dissipative self-force, and the torus-averaged second-order dissipative self-force~\cite{FlanaganHinderer1}. While there are different approximations and ways to construct EMRI waveforms~\cite{Barack:2003fp,Babak:2006uv,Sopuerta:2011te,Chua:2017ujo,Keeble:2025wnp}, state-of-the-art adiabatic EMRI waveforms are built from solutions to the frequency-domain Teukolsky equation~\cite{Hughes_2021,Katz_2021,Chua:2021Rapid,Chapman_Bird_2025}. Waveform modeling sufficiently accurate for parameter estimation may require the inclusion of post-adiabatic corrections, including second-order self-force effects~\cite{Pound_2021, Burko_2013, FlanaganHinderer1, Isoyama_2013} and spin of the secondary~\cite{Drummond:2026haw,Drummond:2023wqc, Skoupy:2023lih, Drummond:2022efc, Mathews:2025txc,Skoupy:2024uan,Piovano:2024yks,Mathews:2025nyb,Drummond:2022xej}. Environmental perturbations, including tidal resonances induced
by a tertiary body~\cite{Gupta:2021cno,Gupta:2022,Bronicki:2022eqa,Cocco:2026lkr,Bonga:2019ycj,Bonga:2026mmz,Destounis:2026tet,Silva:2022blb,Silva:2025lkl}, may also need to be modeled.

The focus of this work is the post-adiabatic effect due to transient orbital resonances. These occur when the radial ($\Omega_{r}$) and polar ($\Omega_{\theta}$) fundamental orbital frequencies (conjugate to Boyer--Lindquist coordinate time) become commensurate~\cite{Contopoulos,FlanaganHinderer2}, i.e., $\beta_r \, \Omega_{\theta} - \beta_{\theta} \, \Omega_{r} = 0$ with $\beta_{r,\theta}\in \mathbb{Z}^{+}$. We refer to such resonances as $\beta_{\theta}{:}\beta_r$. Between capture, inspiral, and plunge, the small body in an EMRI will, in principle, pass through an infinite number of these resonances~\cite{Ruangsri:2013hra}. If neglected, an individual resonance may induce an accumulated GW phase error $\Delta\Phi\sim1/\sqrt{\nu\,\eta}$, where $\nu=\ab{\beta_{r}}+\ab{\beta_{\theta}}$ denotes the order of the resonance~\cite{FlanaganHinderer2}. Low-order resonances in particular may therefore pose a challenge to the GW phase accuracy requirements of EMRI waveform modeling for parameter estimation. 

The analytical treatment and phenomenology of orbital resonances have been studied extensively, e.g., in Refs.~\cite{FHR,FlanaganHinderer2,Ruangsri:2013hra,vandeMeent:2013sza,Brink:2013nna,Brink:2015roa,Isoyama:2013yor,Lynch:2024ohd,Nasipak:2021,Nasipak:2022,Levati,Canizares:2026tgy}. Complementing these theoretical developments, several studies have explored the implications of resonant dynamics for gravitational-wave data analysis. For instance, for an astrophysically motivated population with low eccentricities, Ref.~\cite{Berry} showed that neglecting orbital resonances is unlikely to affect EMRI detection, as the associated loss of signal-to-noise ratio is modest. Other studies have shown that neglecting these resonances can introduce significant systematic biases in parameter estimation~\cite{SperiGair,Levati:2026emw}, demonstrating the importance of resonance modeling for unlocking the full scientific potential of EMRI sources. 

In this work, we revisit the theoretical aspects and phenomenology of orbital resonances through a systematic study of Teukolsky-based resonant flux modifications for a point particle held on a geodesic worldline in Kerr. We analytically derive a selection rule requiring \textit{at least} $\mathrm{lcm}(2,\beta_r)$ cycles in the flux modifications over $\chi_0\in[0,2\pi)$, where $\mathrm{lcm}$ denotes the least common multiple. Associated with this selection rule is a suppression of resonances with odd $\beta_{r}$, which can compete with the degree of $2$-torus sampling as $\beta_{r,\theta}$ increase to determine the relative strengths of the flux modification for different orbital resonances. We argue that the $3{:}2$ generically being the most dynamically relevant --- as observed here and in other works~\cite{FHR,Berry} --- is a natural consequence of optimally balancing these competing effects.

Using the publicly available \texttt{pybhpt} code~\cite{Nasipak:2022,Nasipak:2025tby,Nasipak_pybhpt_2026} to solve the frequency-domain Teukolsky equation, we investigate the $3{:}2$ and $2{:}1$ resonances in detail, examining their dependence on black-hole spin, orbital eccentricity, inclination, and the radial-polar phase $\chi_0$. We also provide results for selected orbital configurations corresponding to the $5{:}2$, $3{:}1$, $5{:}4$, and $4{:}3$ resonances. For the $3{:}2$ and $2{:}1$ resonances, the flux modifications on average increase monotonically with spin and eccentricity. For some orbital configurations, however, the angular-momentum and Carter-constant modifications can be strongly suppressed by destructive interference between the horizon and infinity contributions. Such cancellations create local regions in parameter space in which the resonance strengths can be re-ordered, e.g., the $2{:}1$ can be stronger than the $3{:}2$ in such cases.

To our knowledge, these calculations constitute the largest set of Teukolsky-based resonance coefficients obtained to date. Together with the analytical results, they provide a theoretical foundation and quantitative input for incorporating transient orbital resonances into EMRI waveform models and assessing their consequences for gravitational-wave data analysis.

The rest of this paper is organized as follows. We provide a brief review of the action-angle variable formalism in the Kerr spacetime in Sec.~\ref{sec:action_angle}. We then review in Sec.~\ref{sec:resonant_flux} the frequency-domain Teukolsky formalism for computing non-resonant and resonant fluxes of the energy, axial angular momentum and Carter constant. In Sec.~\ref{sec:results}, we characterize the resonance-induced flux modifications, investigating their phase dependence, relative strengths, and dependence on the primary BH spin, orbital eccentricity and inclination. We discuss the implications of our findings and outline directions for future work in Sec.~\ref{sec:conclusions}. In Appendix~\ref{sec:appendix_A}, we derive the selection rule which determines the minimum number of cycles of the resonant flux modifications over $\chi_{0}\in[0,2\pi)$. In Appendix~\ref{sec:appendix_B}, we describe the convergence scheme adopted for the resonant mode sums and validate our numerical implementation against reference calculations. Throughout, we work in units in which $G=c=1$.

\section{Action-angle variables in the Kerr spacetime}
\label{sec:action_angle}
The orbital dynamics of EMRIs admit a two-timescale expansion~\cite{Mino2005, FlanaganHinderer1, MillerPound2021}: the rapid motion of the secondary object around the central supermassive BH occurs on orbital timescales $t_{\text{orb}}\sim M$, while the inspiraling motion is driven by the dissipation of orbital energy and angular momentum on radiation-reaction timescales $t_{\text{rr}} \sim M/\eta$. The orbital decay occurs due to gravitational back-reaction of the secondary object on its worldline. This gravitational self-force can be split into conservative and dissipative parts. The former is responsible for shifting the orbital frequencies away from their geodesic values, while the latter drives the small body's inspiral~\cite{GSF,Pound_2021,Ruangsri:2013hra}.

Using the relativistic generalization of \textit{action-angle} variables for bound orbits in the Kerr spacetime~\cite{Schmidt:2002qk, Glampedakis:2005cf, FlanaganHinderer1,Pound_2021}, one can expand the self-acceleration of the smaller object in powers of $\eta\ll1$. The equations governing the self-forced motion of the secondary can be cast in terms of action variables $J_{i}$ and angle variables $q_{\alpha}$ to order $\order{\eta^2}$ as~\cite{FlanaganHinderer1}
\begin{align}
    & \dv{q_{\alpha}}{\tau} = \omega_{\alpha} (J_k) + \eta g_{\alpha}^{(1)}(q_A, J_k) + \eta^2 g_{\alpha}^{(2)}(q_A, J_k), \label{angle} \\
    & \dv{J_{i}}{\tau} = \eta G_{i}^{(1)}(q_A, J_k) + \eta^2 G_{i}^{(2)}(q_A, J_k), \label{action}
\end{align}
where $q_\alpha\in\{q_{t}, q_{r},q_{\theta},q_{\phi}\}$, $q_{A}\in\{q_{r},q_{\theta}\}$, $i,k\in\{1,2,3\}$ and $\tau$ is the proper time along the secondary's worldline. The functions $\omega_{i} (J_k)$ are the fundamental radial, polar and azimuthal frequencies conjugate to $\tau$. We denote by $\Omega_{i}$ the corresponding frequencies conjugate to
Boyer--Lindquist coordinate time $t$. The forcing terms $g_{\alpha}^{(1)}$, $G_{i}^{(1)}$ and $g_{\alpha}^{(2)}$, $G_{i}^{(2)}$ denote first- and second-order pieces of the gravitational self-force (GSF), respectively.

At zeroth order in $\eta$, Eqs.~(\ref{angle}--\ref{action}) reduce to bound Kerr geodesic motion with conserved quantities $J_i\in\{E,L_z,Q\}$, where $E$ is the orbital energy, $L_z$ is azimuthal angular momentum and $Q$ is the Carter constant. These constants can be mapped to geometric orbital elements through the following parametrization of $r$ and $\theta$~\cite{Schmidt:2002qk, Fujita:2009us}
\begin{align}
    & r = \frac{p}{1 + e \cos \psi}, \hspace{2cm} r_p \leq r \leq r_a, \label{parameterizationR} \\
    & \cos\theta = \cos\theta_{\min} \cos\chi, \hspace{1cm} \theta_{\min} \leq \theta \leq \pi - \theta_{\min},
\end{align}
where $r_p$ is the periapsis, $r_a$ is the apoapsis, $p$ is the semi-latus rectum and $e$ is the eccentricity. The radial turning points are related to $p$ and $e$ by $r_p = p / (1 + e)$ and $r_a = p / (1 - e)$. The initial conditions of the orbit are specified by $\psi(t=0)\equiv\psi_{0}$ and $\chi(t=0)\equiv\chi_{0}$. Throughout, we make use of the inclination parameter $x_I=\cos I$, where $I=\pi/2-\mathrm{sgn}(L_{z})\theta_{\min}$ is the inclination angle relative to the equatorial plane. In this work, we focus only on prograde orbits and therefore take $\mathrm{sgn}(L_{z})=1$. Mappings between the parametrizations $(p, e, x_I)$ and $(E, L_z, Q)$ are provided in Refs.~\cite{Schmidt:2002qk,Hughes2024}. 

At higher orders in $\eta$, self-force corrections drive the motion away from the geodesic worldline. Off resonance, these higher-order corrections enter hierarchically: the $2$-torus (i.e., $q_{r}$--$q_{\theta}$) average of the first-order dissipative self-force drives the leading adiabatic contribution to the orbital phase at $\mathcal{O}(1/\eta)$, while the oscillatory first-order self-force and $2$-torus average of the dissipative second-order self-force provide the sub-leading post-$1$-adiabatic correction at $\mathcal{O}(1)$~\cite{FlanaganHinderer1}. The adiabatic approximation consists of dropping the phase-forcing term $g_\alpha^{(1)}$ in Eq.~\eqref{angle} and replacing the action-forcing term $G_i^{(1)}$ in Eq.~\eqref{action} by its orbit-averaged dissipative contribution. In this approximation, $G_i^{(1)}$ goes to $\langle G_i^{(1)}\rangle_{q_{r}q_{\theta}}$, which drives the adiabatic inspiral~\cite{FlanaganHinderer1}. Combined with the osculating orbits approach~\cite{PoundPoisson:2008Osculating,Pound_2021}, the adiabatic EMRI inspiral is constructed as a flow through a sequence of geodesic orbits, each one with updated values of $(E,L_z,Q)$ from the adiabatic fluxes $\langle \dot J_i\rangle$ to account for radiation reaction. 

At an orbital resonance, a sub-leading correction to the phase evolution is introduced at $\mathcal{O}(1/\sqrt{\eta})$, in addition to a phase-dependent change to the leading-order orbit-averaged forcing~\cite{FlanaganHinderer2}. Fourier expanding $G_{i}^{(1)}(q_{r},q_{\theta},J_{k})$ to first order in $\eta$, the $2$-torus average of Eq.~\eqref{action} is given by
\begin{equation}
\begin{split}
\left\langle \frac{\ed J_i}{\ed \tau}\right\rangle
= \,
& \eta \, G_{i,00}(J) \, +
\\
&
\eta\sum_{(k,n)\neq(0,0)}
G_{i,kn}(J)
\left\langle e^{i(k q_\theta - n q_r)}\right\rangle_{q_\theta q_r},
\end{split}
\label{eq:fourier_flux}
\end{equation}
where $k$ and $n$ denote the polar and radial harmonic indices, respectively. Off resonance, the post-adiabatic $(k,n)\neq(0,0)$ corrections in Eq.~\eqref{eq:fourier_flux} are highly oscillatory and average to zero over many orbital cycles, leaving only the adiabatic term $G_{i,00}$ to govern the secular inspiral. As the inspiral proceeds, radiation reaction slowly changes fundamental frequencies and the system will, in principle, pass through an infinite number of transient orbital resonances~\cite{Ruangsri:2013hra}, whereupon the fundamental frequencies become commensurate. At a $\beta_{\theta}{:}\beta_{r}$ resonance, the $(k,n)=(\beta_{r}, \beta_{\theta})$ term in Eq.~\eqref{eq:fourier_flux} is slowly varying and survives the orbit average, providing a transient ``kick'' to the constants of motion~\cite{FlanaganHinderer2}. The kick arises as a result of self-force contributions that ordinarily average to zero over many orbital cycles instead interfering coherently over the resonance timescale, introducing additional, phase-dependent contributions to the fluxes~\cite{FlanaganHinderer2,FHR}. The resonant kick causes the trajectory to depart from the adiabatic evolution, leading to a corresponding dephasing of the gravitational waveform if neglected~\cite{vandeMeent:2013sza,FlanaganHinderer2,FHR}.

\section{Resonant Flux Formalism}
\label{sec:resonant_flux}

A Teukolsky-based prescription for computing the resonance-induced modifications to the fluxes in the strong-field regime was developed by Flanagan, Hughes and Ruangsri~\cite{FHR}. The precise value of the fluxes on resonance varies with the relative phase of the radial and polar motions~\cite{FlanaganHinderer2, FHR, Berry}: two orbits with the same values of ($E$, $L_z$, $Q$) which enter the resonance with different phases can receive different resonant kicks, changing the subsequent evolution of the orbits. In this way, resonances enhance the dependence of the EMRI evolution on initial conditions~\cite{FlanaganHinderer2}.

In this section, we briefly review the computation of the resonant and non-resonant fluxes of $(E,L_{z},Q)$ for a point particle held on a geodesic worldline using the frequency-domain Teukolsky equation based on the work of Ref.~\cite{FHR}. Following Ref.~\cite{FHR}, we choose the origin of time such that the orbit is at periapsis, i.e., $\psi_0=0$. The remaining phase offset $\chi_0$ specifies the polar phase at periapsis and therefore determines the relative phase between the radial and polar motions. The fiducial geodesic is defined by additionally setting $\chi_0=0$.

\subsection{Non-resonant fluxes}

First-order gravitational perturbations of the Kerr spacetime are described by the spin-weight $s=-2$ Teukolsky equation~\cite{TeukolskyPress1974}, which takes the schematic form $\mathcal{D}\br{\psi_4}=\mathcal{T},$ where $\psi_{4}$ is a complex curvature scalar encoding the two gravitational degrees of freedom at each point in spacetime, $\mathcal{D}$ is a second-order linear differential operator, and $\mathcal{T}$ is a source term~\cite{1973ApJ...185..635T}. The ansatz
\begin{align}
    \rho^{-4}\psi_{4}=\int_{-\infty}^{\infty}\ed \omega\sum_{\ell,m}R_{\ell m\omega}(r)S_{\ell m\omega}(\theta)e^{-i\omega t+im\phi},
\end{align}
where $\rho=-1/\pa{r-ia\cos{\theta}}$, separates the Teukolsky equation into an angular equation for the spin-weighted spheroidal harmonics $S_{\ell m\omega}(\theta)$ and a radial equation
\begin{align}
    \Delta^{2}\frac{\ed}{\ed r}\pa{\frac{1}{\Delta}\frac{\ed R_{\ell m\omega}}{\ed r}}-V(r)R_{\ell m\omega}=-\mathcal{T}_{ \ell m\omega}(r; \chi_{0}),\label{eq:radial}
\end{align}
where $\Delta=r^2-2Mr+a^2$. Hereafter, we refer to Eq.~\eqref{eq:radial} as the Teukolsky equation.

A convenient basis of homogeneous solutions to the Teukolsky equation consists of the so-called ``in'' and ``up'' modes. The in-mode is purely ingoing at the future horizon and consists of incident and reflected amplitudes at past and future null infinity, respectively, while the up-mode is the complementary solution which is purely outgoing at future null infinity~\cite{TeukolskyPress1974,ManoSuzukiTakasugi1996,SasakiTagoshi2003}. A particular solution to the Teukolsky equation is given by convolution of the in- and up-modes with the source $\mathcal{T}_{\ell m\omega}$ according to the method of Green's functions. Asymptotically, a particular solution takes the form
\begin{align}
    R_{\ell m\omega}(r)=
    \begin{cases}
     Z^{\rm{up}}_{\ell m\omega}(\chi_{0})R^{\rm{in}}_{\ell m\omega}(r)\quad &r\to r_{+},\\
     Z^{\rm{in}}_{\ell m\omega}(\chi_{0})R^{\rm{up}}_{\ell m\omega}(r)\quad &r\to \infty,
    \end{cases}
\end{align}
where $r_+=M+\sqrt{M^2-a^2}$, 
\begin{align}
    Z^{\star}_{\ell m\omega}=C^{*}_{\ell m\omega}\int_{r_{+}}^{\infty}\frac{R^{\star}_{\ell m\omega}(\xi)\mathcal{T}_{\ell m\omega}(\xi;\chi_{0})}{\Delta(\xi)^{2}}\ed \xi,\label{eq:Z1}
\end{align}
with the symbol $\star\in\{\rm{in},\rm{up}\}$ and $C^{\star}_{\ell m\omega}$ is a constant, related to the Wronskian of the homogeneous solutions. 

The source $\mathcal{T}_{\ell m\omega}$ is constructed from the stress-energy tensor of a point particle undergoing bound geodesic motion in the background Kerr spacetime~\cite{Drasco:2005kz, FHR}. Consequently, it has support only on the worldline of the geodesic and at the discrete harmonics of the fundamental frequencies
\begin{equation}
    \omega_{mkn}
    =
    m\Omega_\phi + k\Omega_\theta + n\Omega_r.
\end{equation}
Inserting the explicit form of the source $\mathcal{T}_{\ell m\omega}$ into Eq.~\eqref{eq:Z1}, converting integrals with respect to proper time to Mino time $\lambda$~\cite{Mino} and Fourier expanding with respect to $\lambda$, one finds~\cite{Drasco:2003ky,FHR}
\begin{align}
    Z^{\star}_{\ell m\omega}=\sum_{k,n}Z^{\star}_{\ell mkn}(\chi_{0})\delta(\omega-\omega_{mkn}),
\end{align}
where $Z^{\star}_{\ell mkn}$ is proportional to an integral over decoupled cycles in $r(\lambda)$ and $\theta(\lambda)$ of a frequency-domain quantity built from the source $\mathcal{T}_{\ell m\omega}$ (see, for instance, Ref.~\cite{FHR} and references therein for further details). The mode amplitudes $Z^{\mathrm{in}}_{\ell mkn}$ and $Z^{\mathrm{up}}_{\ell mkn}$ determine the fluxes of $E$, $L_z$ and $Q$ carried to future null infinity and across the future event horizon, respectively.

We define the fiducial amplitudes as $\check{Z}^{\star}_{\ell mkn}\equiv Z^{\star}_{\ell mkn}(\chi_{0}=0)$. The dependence of the amplitudes $Z^{\star}_{\ell mkn}$ on the relative radial-polar phase $\chi_0$ can be isolated into a phase factor multiplying the fiducial amplitudes,
\begin{equation}
    Z^\star_{\ell mkn}(\chi_0)
    =
    e^{i\xi_{mkn}(\chi_0)}
    \check{Z}^\star_{\ell mkn}.
\end{equation}
The phase shift $\xi_{mkn}(\chi_0)$ is constructed from the oscillatory contributions to $t(\lambda)$ and $\phi(\lambda)$ along the fiducial orbit and it satisfies $\xi_{mkn}(0)=0$. It is defined in Eq.~$(3.18)$ of Ref.~\cite{FHR} as
\begin{align}
\xi_{mkn}(\chi_0)
&= k\Upsilon_\theta \lambda_0^\theta
+ m\Delta\hat{\phi}\br{r_{\mathrm{min}},\theta(-\lambda_0^\theta)} \nonumber \\
&\quad - \omega_{mkn}\,\Delta\hat{t}\br{r_{\mathrm{min}},\theta(-\lambda_0^\theta)},
\label{eq:phase}
\end{align}
where $\lambda_0^\theta = \lambda_0^\theta(\chi_0)$ is the value of a Mino-time variable at which $\theta=\theta_{\min}$, $\Upsilon_{\theta}$ is the polar frequency conjugate to Mino time, and $\Upsilon_\theta\lambda_0^\theta(\chi_0)\equiv q_{\theta0}(\chi_0)$. 

For a generic, non-resonant orbit, the long-time average eliminates cross terms between modes with different frequencies  $\omega_{mkn}$, forcing $m=m'$, $k=k'$ and $n=n'$ in Eq.~(3.23) of Ref.~\cite{FHR} and leaving only diagonal
contributions to the fluxes proportional to
\begin{equation}
    \left|Z^\star_{\ell mkn}(\chi_0)\right|^2
    =
    \left|
    e^{i\xi_{mkn}(\chi_0)}
    \check{Z}^\star_{\ell mkn}
    \right|^2
    =
    \left|\check{Z}^\star_{\ell mkn}\right|^2,
\end{equation}
which have no dependence on initial relative phase $\chi_0$. The resulting energy and axial angular momentum fluxes to infinity and across the horizon are given by~\cite{FHR}
\begin{align}
    \left\langle\frac{\ed E^\infty}{\ed t}\right\rangle
    &=
    \sum_{\ell mkn}
    \frac{|\check{Z}^{\mathrm{in}}_{\ell mkn}|^2}
         {4\pi\omega_{mkn}^2},
    \label{eq:Eflux_nonres}
    \\
    \left\langle\frac{\ed E^H}{\ed t}\right\rangle
    &=
    \sum_{\ell mkn}
    \alpha_{\ell mkn}
    \frac{|\check{Z}^{\mathrm{up}}_{\ell mkn}|^2}
         {4\pi\omega_{mkn}^2},
\end{align}
and
\begin{align}
     \left\langle\frac{\ed L_z^\infty}{\ed t}\right\rangle
    &=
    \sum_{\ell mkn}
    \frac{m|\check{Z}^{\mathrm{in}}_{\ell mkn}|^2}
         {4\pi\omega_{mkn}^3},
    \label{eq:Lzflux_nonres}
    \\
    \left\langle\frac{\ed L_z^H}{\ed t}\right\rangle
    &=
    \sum_{\ell mkn}
    \alpha_{\ell mkn}
    \frac{m|\check{Z}^{\mathrm{up}}_{\ell mkn}|^2}
         {4\pi\omega_{mkn}^3},
\end{align}
where $\alpha_{\ell mkn}$ accounts for the absorption of radiation by the
black-hole horizon. 

Unlike $E$ and $L_z$, the Carter constant $Q$ is not associated with a flux that can be directly extracted from the gravitational radiation at infinity or at the black-hole horizon. Its secular evolution due to radiative backreaction can be computed taking into account only the dissipative part of the GSF and averaging over long times. The corresponding expressions are~\cite{Sago:2005RadiationReaction,Sago:2005fn,FHR}
\begin{align}
    \left\langle\frac{\ed Q^\infty}{\ed t}\right\rangle
    &=
    \sum_{\ell mkn}
    |\check{Z}^{\mathrm{in}}_{\ell mkn}|^2
    \frac{\mathcal{L}_{mkn}+k\Upsilon_\theta}
         {2\pi\omega_{mkn}^3},
    \\
    \left\langle\frac{\ed Q^H}{\ed t}\right\rangle
    &=
    \sum_{\ell mkn}
    \alpha_{\ell mkn}|\check{Z}^{\mathrm{up}}_{\ell mkn}|^2
    \frac{\mathcal{L}_{mkn}+k\Upsilon_\theta}
         {2\pi\omega_{mkn}^3},
\end{align}
where
\begin{equation}
    \mathcal{L}_{mkn}
    =
    m\langle\cot^2\theta\rangle L_z
    -
    a^2\omega_{mkn}
    \langle\cos^2\theta\rangle E .
\end{equation}

\subsection{Resonant fluxes}

On resonance, it is useful to introduce the resonant frequency
\begin{equation}
    \Omega_{\rm res}
    \equiv
    \frac{\Omega_\theta}{\beta_\theta}
    =
    \frac{\Omega_r}{\beta_r}.
\end{equation}
The radial and polar contributions to the mode frequency $\omega_{mkn}$ can then be
combined as
\begin{equation}
    k\Omega_\theta+n\Omega_r
    =
    \left(k\beta_\theta+n\beta_r\right)\Omega_{\rm res}
    =
    N\Omega_{\rm res},
\end{equation}
where
\begin{equation}
    N\equiv k\beta_\theta+n\beta_r.
\end{equation}
Consequently, the mode frequency depends on the polar and radial harmonic indices $(k,n)$ only through the single integer $N$ and can thus be written as
\begin{equation}
    \omega_{mN}
    =
    m\Omega_\phi+N\Omega_{\rm res}.\label{eq:omegamN}
\end{equation}

For a fixed pair $(m,N)$, an infinite number of pairs $(k,n)$ correspond to the same frequency $\omega_{mN}$. Given one pair $(k_0,n_0)$ such that
\begin{equation}
    k_0\beta_\theta+n_0\beta_r=N,
\end{equation}
all members of the resonant family can be written as
\begin{equation}
    k_j = k_0+j\beta_r,
    \qquad
    n_j = n_0-j\beta_\theta,
    \qquad j\in\mathbb{Z},
    \label{eq:res_family}
\end{equation}
since 
\begin{equation}
    k_j\beta_\theta+n_j\beta_r
    =
    k_0\beta_\theta+n_0\beta_r
    =
    N.
\end{equation}
Thus, each $(\ell,m,N)$ mode contains an entire family of $(k_j,n_j)$ harmonics which are degenerate in frequency and whose associated amplitudes interfere on resonance. This behavior is in contrast to generic orbits, for which modes with distinct $(k,n)$ have distinct frequencies and do not interfere due to their cross
terms vanishing under the long-time average. 

The coherent resonant amplitude is
\begin{equation}
    Z^\star_{\ell mN}(\chi_0)
    =
    \sum_{(k,n)_N}
    e^{i\xi_{mkn}(\chi_0)}
    \check{Z}^\star_{\ell mkn},
    \label{eq:resonant_Z}
\end{equation}
where $(k,n)_N$ denotes all pairs satisfying
$k\beta_\theta+n\beta_r=N$. Using the explicit
parametrization of the resonant family introduced in
Eq.~\eqref{eq:res_family}, Eq.~\eqref{eq:resonant_Z} becomes
\begin{align}
    Z^\star_{\ell mN}(\chi_0)
    &=
    \sum_{j=-\infty}^{\infty}
    e^{i\xi_{m k_j n_j}(\chi_0)}
    \check{Z}^\star_{\ell m k_j n_j},
\label{eq:resonant_Z_j}
\end{align}
which, crucially, has phase factors
$e^{i\xi_{m k_j n_j}(\chi_0)}$ appearing inside the sum over $j$.
Consequently, the quantity
\begin{equation}
    \left|Z^\star_{\ell mN}(\chi_0)\right|^2
    =
    \left|
        \sum_{j=-\infty}^{\infty}
        e^{i\xi_{m k_j n_j}(\chi_0)}
        \check{Z}^\star_{\ell m k_j n_j}
    \right|^2 
\label{eq:resonant_Z_j_squared}
\end{equation}
contains cross terms between different members of the resonant family. These cross terms depend on the relative phases
$\xi_{m k_j n_j}(\chi_0)-\xi_{m k_{j'} n_{j'}}(\chi_0)$. Using Eq.~\eqref{eq:phase}, the phase difference between two members $(k,n)$ and $(k',n')$ of the same resonant family is given by
\begin{equation}
    \xi_{mkn}(\chi_0)-\xi_{mk'n'}(\chi_0)
    =
    (k-k')q_{\theta0}(\chi_0),
\label{eq:delta_phase_factor}
\end{equation}
as shown in Eq.~(4.3) of Ref.~\cite{FHR}. The phase difference \eqref{eq:delta_phase_factor} between different members of the same resonant family controls whether they interfere constructively or destructively.

The resonant energy and axial angular momentum fluxes to infinity and across the horizon are given by~\cite{FHR}
\begin{align}
\left\langle
\frac{\ed E^\infty}{\ed t}(\chi_0)
\right\rangle
&=
\sum_{\ell mN}
\frac{|Z^{\mathrm{in}}_{\ell mN}(\chi_0)|^2}
     {4\pi\omega_{mN}^2}\nonumber\\
&\equiv
\sum_{\ell mN}
\dot{E}^\infty_{\ell mN}(\chi_0),
\label{eq:Einf_res}
\\
\left\langle
\frac{\ed E^H}{\ed t}(\chi_0)
\right\rangle
&=
\sum_{\ell mN}
\alpha_{\ell mN}
\frac{|Z^{\mathrm{up}}_{\ell mN}(\chi_0)|^2}
     {4\pi\omega_{mN}^2}\nonumber\\
&\equiv
\sum_{\ell mN}
\dot{E}^H_{\ell mN}(\chi_0),
\label{eq:EH_res}
\end{align}
and
\begin{align}
\left\langle
\frac{\ed L_z^\infty}{\ed t}(\chi_0)
\right\rangle
&=
\sum_{\ell mN}
\frac{m|Z^{\mathrm{in}}_{\ell mN}(\chi_0)|^2}
     {4\pi\omega_{mN}^3}\nonumber\\
&\equiv
\sum_{\ell mN}
\dot{L}_{z,\ell mN}^\infty(\chi_0),
\label{eq:Lzinf_res}
\\
\left\langle
\frac{\ed L_z^H}{\ed t}(\chi_0)
\right\rangle
&=
\sum_{\ell mN}
\alpha_{\ell mN}
\frac{m|Z^{\mathrm{up}}_{\ell mN}(\chi_0)|^2}
     {4\pi\omega_{mN}^3}\nonumber\\
&\equiv
\sum_{\ell mN}
\dot{L}_{z,\ell mN}^H(\chi_0).
\label{eq:LzH_res}
\end{align}
The resonant Carter-constant fluxes are given by~\cite{FHR}
\begin{align}
\left\langle
\frac{\ed Q^\infty}{\ed t}(\chi_0)
\right\rangle
&=\Upsilon_\theta
\sum_{\ell mN}
\frac{
\Re\!\left[
Z^{\mathrm{in}}_{\ell mN}(\chi_0)
\overline{Y^{\mathrm{in}}_{\ell mN}(\chi_0)}
\right]}
{2\pi\omega_{mN}^3}
\nonumber\\
&\quad+\sum_{\ell mN}
\frac{|Z^{\mathrm{in}}_{\ell mN}(\chi_0)|^2}
     {2\pi\omega_{mN}^3}
\mathcal{L}_{mN},
\label{eq:Qinf_res}
\\[1ex]
\left\langle
\frac{\ed Q^H}{\ed t}(\chi_0)
\right\rangle
&=\Upsilon_\theta
\sum_{\ell mN}
\alpha_{\ell mN}
\frac{
\Re\!\left[
Z^{\mathrm{up}}_{\ell mN}(\chi_0)
\overline{Y^{\mathrm{up}}_{\ell mN}(\chi_0)}
\right]}
{2\pi\omega_{mN}^3}
\nonumber\\
&\quad+\sum_{\ell mN}
\alpha_{\ell mN}
\frac{|Z^{\mathrm{up}}_{\ell mN}(\chi_0)|^2}
     {2\pi\omega_{mN}^3}
\mathcal{L}_{mN},
\label{eq:QH_res}
\end{align}
where
\begin{equation}
Y^\star_{\ell mN}(\chi_0)
=
\sum_{(k,n)_N}
k\,e^{i\xi_{mkn}(\chi_0)}
\check{Z}^\star_{\ell mkn},
\label{eq:resonant_Y}
\end{equation}
and the quantity $\mathcal{L}_{mN}$ is obtained from the corresponding
non-resonant quantity $\mathcal{L}_{mkn}$ by replacing
$\omega_{mkn}$ with $\omega_{mN}$. The fluxes in Eqs.~\eqref{eq:Qinf_res} and~\eqref{eq:QH_res} are purely dissipative and do not include the additional conservative self-force contribution identified in Ref.~\cite{Isoyama:2019FluxBalance}, for which numerical evidence was presented in a scalar model in Ref.~\cite{Nasipak:2022}.

The total resonant flux is obtained by summing the
individual $(\ell,m,N)$ mode contributions,
\begin{equation}
    \dot{J}^{\star}(\chi_0)
    =
    \sum_{\ell=2}^{\infty}
    \sum_{m=-\ell}^{\ell}
    \sum_{N=-\infty}^{\infty}
    \dot{J}^{\star}_{\ell mN} (\chi_0).
    \label{eq:total_res_flux}
\end{equation}
To summarize, the resonant flux calculation involves two successive summations. For each fixed $(\ell,m,N)$, the degenerate $(k_j,n_j)$ harmonics within the same resonant family are first combined coherently at the amplitude level to compute the mode contribution $\dot{J}^{\star}_{\ell mN}$. These contributions are then summed over $(\ell,m,N)$ to obtain the total resonant flux.

\begin{figure*}[]
\centering
\includegraphics[width=\textwidth]{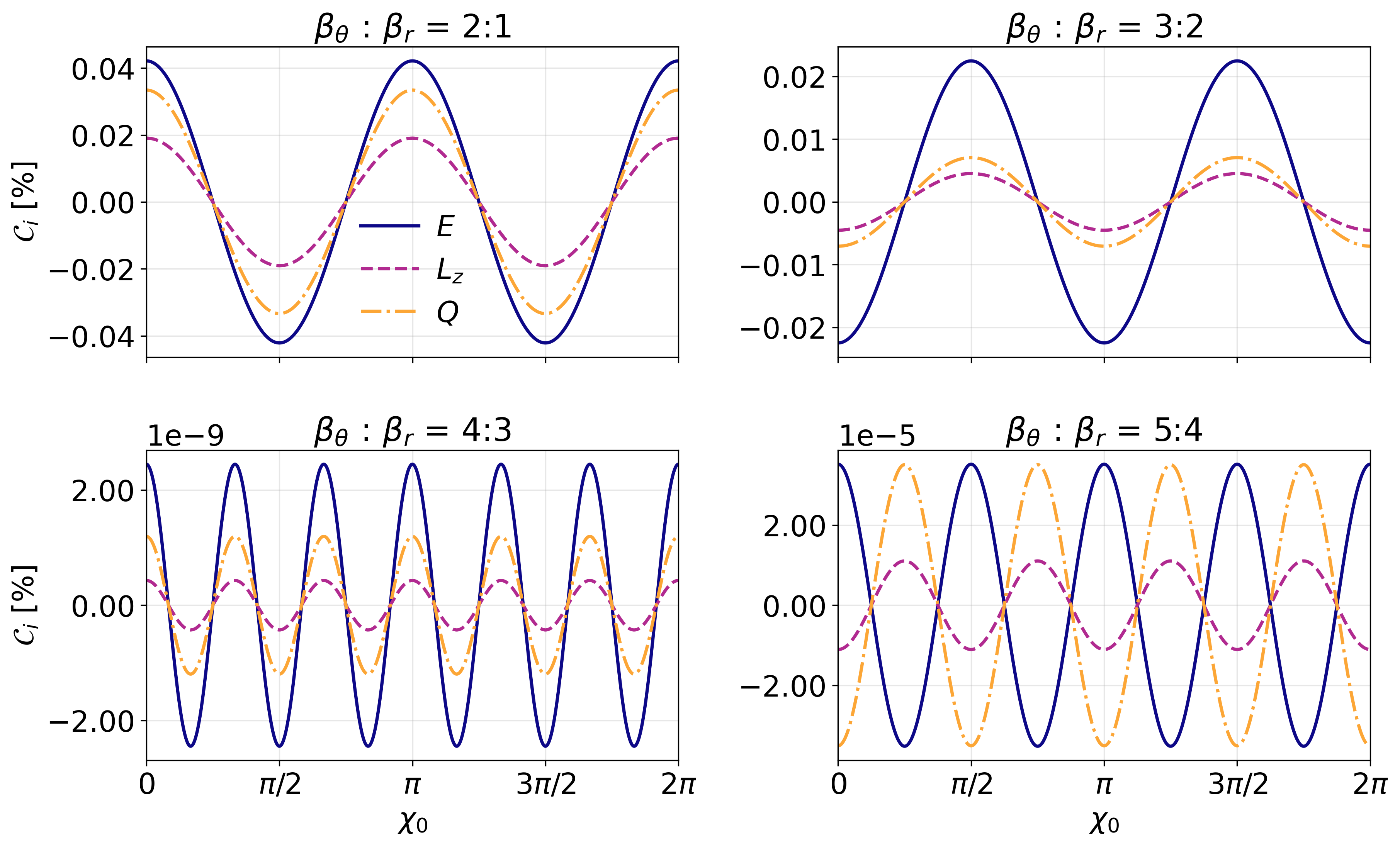}
\caption{Phase dependence of the resonance coefficients for the energy (solid lines), axial angular momentum (dashed lines) and Carter constant (dash-dotted lines) fluxes. The orbital parameters $(a,e,x_I,p/M)$ are $(0.90,0.70,0.93969,3.78947)$, $(0.90,0.30,0.93969, 5.34138)$, $(0.90,0.30,0.34202,11.31578)$ and $(0.90,0.70,0.34202,14.13622)$ for the $2{:}1$, $3{:}2$, $4{:}3$ and $5{:}4$ resonances, which each exhibit $2$, $2$, $6$ and $4$ cycles, respectively,  over $\chi_0\in[0,2\pi)$. These periodicities obey the selection rule $N_{\rm cyc}=\mathrm{lcm}(2,\beta_r)$ as predicted by Eq.~\eqref{eq:EmodeSum}.}
\label{fig:periodicity}
\end{figure*}

\section{Resonance-Driven Flux Modifications}
\label{sec:results}

With all the pieces together, we can now compute the resonance coefficients
\begin{equation}
\mathcal{C}_i (\chi_0) =
\frac{\dot{J_i} (\chi_0) - \langle\dot{J_i}\rangle_{\chi_0}
}{\langle\dot{J_i}\rangle_{\chi_0}},
\label{eq:coeffs}
\end{equation}
where $i\in\{E,L_z,Q\}$ labels the orbital integral. In Eq.~\eqref{eq:coeffs}, $\dot{J_i} (\chi_0)$ denotes the flux evaluated on resonance at a given relative phase $\chi_{0}$, and $\langle\dot{J_i}\rangle_{\chi_0}$ the corresponding non-resonant flux given by its average over $\chi_{0}\in[0,2\pi)$. The coefficients $\mathcal{C}_i$ define normalized phase-dependent flux modifications which are used as input to phenomenological effective resonance models in data analysis~\cite{SperiGair,Levati:2026emw}. Throughout, we compare the ``strength'' of these modifications for different resonances, as encoded by the size of these dimensionless coefficients. These coefficients are not to be confused with jumps in the constants of motion, whose computation requires additional dynamical input~\cite{FlanaganHinderer2,vandeMeent:2013sza}, and which are approximated in effective resonance models. For all calculations, we numerically evaluate the flux formulae in Sec.~\ref{sec:resonant_flux} using \texttt{pybhpt}~\cite{Nasipak:2022,Nasipak:2025tby,Nasipak_pybhpt_2026}.

We now characterize the phase dependence, strength, and parameter-space dependence of the resonance-induced flux modifications. All reported results satisfy the mode-sum and flux-error criteria described in App.~\ref{sec:appendix_B}. Tables~\ref{tab:fhr_comparison_1}--\ref{tab:fhr_comparison_4} therein also provide numerical values of the peak-to-trough variation associated with the horizon, infinity, and total fluxes for the orbits considered in Ref.~\cite{FHR}.

\subsection{Phase dependence}

The resonant flux modifications depend on the relative phase between the radial and polar motions. The periodicity of the modifications with respect to $\chi_{0}$ can be determined from the equations for the fluxes. As an example, we consider the energy flux to infinity as given in Eq.~\eqref{eq:Einf_res}. It follows from Kerr's discrete equatorial symmetry $\theta \to \pi - \theta$ that~\cite{Drasco:2005kz}
\begin{equation}
\check{Z}^{\star}_{\ell,-m,-k,-n}
= (-1)^{\ell+k}\bar{\check{Z}}^{\star}_{\ell m k n},
\label{eq:equatorial_symmetry}
\end{equation}
which allows one to reduce the sum over $(m,N)$ in Eq.~(\ref{eq:Einf_res}) to a sum over $(m>0,N\in\mathbb{Z})$ and $(m=0,N>0)$. The resulting sum consists of resonant contributions of the form $\dot{E}^{\infty,\mathrm{res}}_{\ell mN}(\chi_0)+\dot{E}^{\infty,\mathrm{res}}_{\ell -m-N}(\chi_0)$. We show in Appendix~\ref{sec:appendix_A} that such contributions take the form
\begin{widetext}
    \begin{align}
        \dot{E}^{\infty,\mathrm{res}}_{\ell mN}(\chi_0)+\dot{E}^{\infty,\mathrm{res}}_{\ell -m-N}(\chi_0)&=\sum_{j>j'}
2\br{1+\pa{-1}^{s(j,j')}}\mathrm{Re}\pa{e^{is(j,j')q_{\theta0}(\chi_0)}\,\cZk{\ell m k_j n_j}\,{\bcZk{\ell m k_{j'} n_{j'}}}},\label{eq:EmodeSum}
    \end{align}
\end{widetext}
where $s(j, j') = (j - j')\,\beta_r \in \cu{\pm\beta_r, \pm 2\beta_r, \dots}$. Only even values of $s(j, j')$ contribute to the sum in Eq.~\eqref{eq:EmodeSum}. If $\beta_{r}$ is even, all permitted values of $s(j, j')$ are even and hence all contribute, while if $\beta_{r}$ is odd, all the permissible odd values of $s(j, j')$ cancel in the fluxes. As shown below, this ``odd cancellation'' appears to correspond to a suppression of resonant modifications for odd $\beta_{r}$. 

The periodicity of the sum of terms on the left-hand side of Eq.~\eqref{eq:EmodeSum} is set by the smallest value of $\ab{s(j,j')}$. Thus, the minimum number of cycles of this quantity for a resonance $\beta_\theta{:}\beta_r$ over $\chi_0\in[0,2\pi)$ is
\begin{equation}
N_{\rm cyc} = \mathrm{lcm}(2,\beta_r),
\end{equation}
where $\mathrm{lcm}$ denotes the least common multiple. This phase dependence is independent of the mode $(\ell, m, N)$ and is thus inherited by the total flux \eqref{eq:Einf_res}. Consequently, resonances with even $\beta_r$ have $N_{\rm cyc}=\beta_r$, whereas those with odd $\beta_r$ have $N_{\rm cyc}=2\beta_r$. 

Figure~\ref{fig:periodicity} illustrates the phase dependence of the resonance coefficients for the energy, axial angular momentum and Carter constant fluxes, where the $2{:}1$, $3{:}2$, $4{:}3$ and $5{:}4$ resonances exhibit $2$, $2$, $6$ and $4$ cycles, respectively. Previous calculations of resonant flux modifications~\cite{FHR} and dissipative scalar self-force contributions~\cite{Nasipak:2021, Nasipak:2022} found the same periodicity for the $3{:}2$ and $2{:}1$ resonances.

All the resonances considered in this work admit values of $\chi_0$ for which the $E$, $L_z$ and $Q$ resonance coefficients vanish simultaneously. The occurrence of these zeros reflects the phase periodicity of each resonance, i.e., the number of zeros is equal to $2N_{\rm{cyc}}$. Figure~\ref{fig:periodicity} also illustrates that the relative signs of the $E$, $L_z$ and $Q$ resonance coefficients depend on both the orbital parameters and the resonance considered. Across the configurations explored in this work, we find both common and mixed signs: simultaneously, all three fluxes can be enhanced, all three suppressed, or some enhanced and others suppressed relative to their non-resonant values.

\begin{figure}[]
\centering
\includegraphics[width=\linewidth]{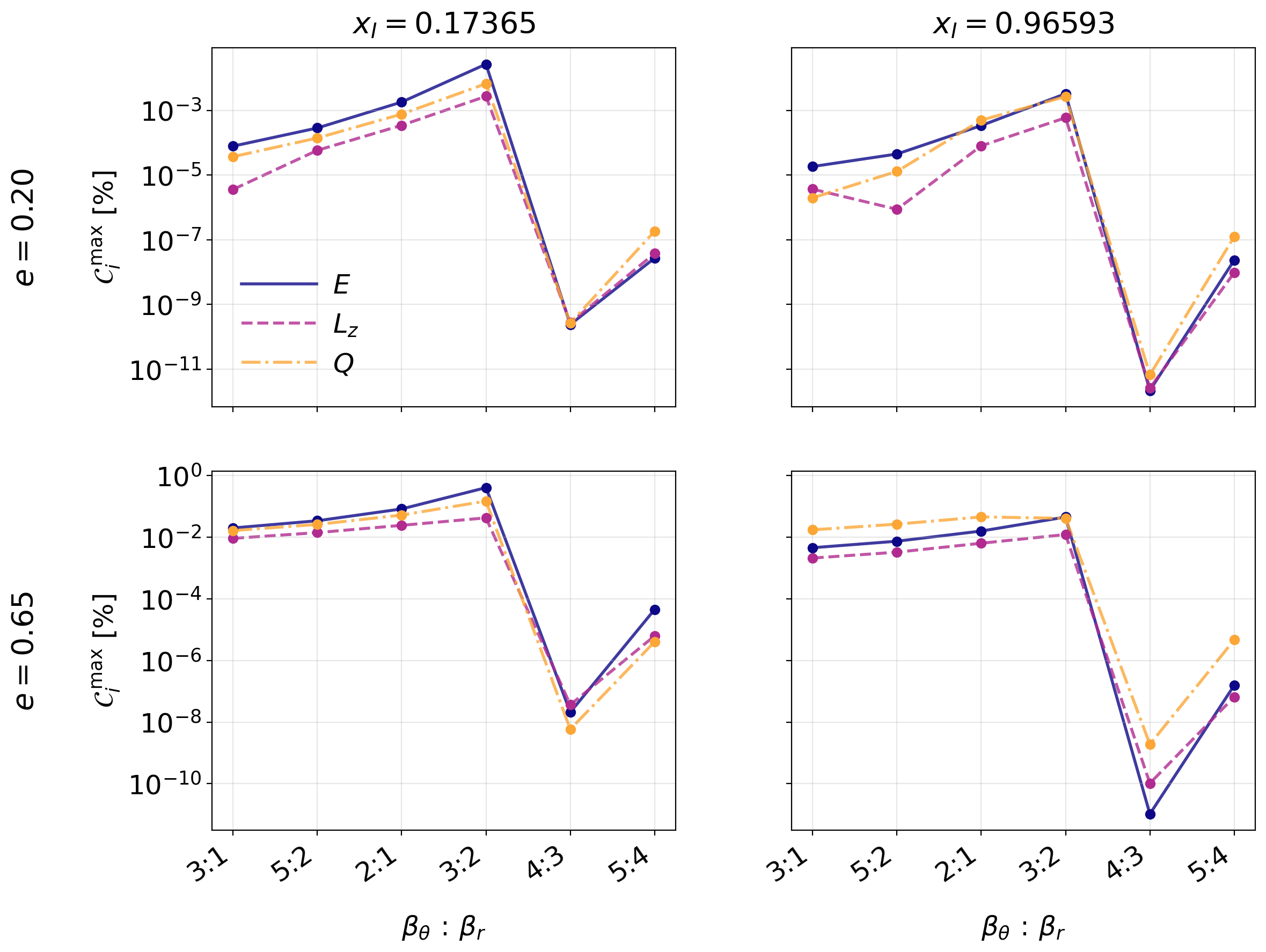}
\caption{Strength of the resonant flux modifications quantified by the maximum resonance coefficients over the phase $\chi_0$, $\mathcal{C}_i^{\rm max}\equiv\max_{\chi_0}|\mathcal{C}_i(\chi_0)|$, for the energy (solid lines), axial angular momentum (dashed lines) and Carter constant (dash-dotted lines) fluxes. The $3{:}1$, $5{:}2$, $2{:}1$, $3{:}2$, $4{:}3$ and $5{:}4$ resonances are ordered by increasing semi-latus rectum. Results are shown for a primary black hole spin $a=0.80$, eccentricities $e=0.20$ and $e=0.65$, and inclinations $x_I=0.17365$ and $x_I=0.96593$.}
\label{fig:res_strength}
\end{figure}

\subsection{Strength of the resonant flux modifications}

We next characterize the strength of the resonant flux modifications by reporting the maximum value of the resonance coefficients over the phase $\chi_0$, i.e., $\mathcal{C}_i^{\rm max} \equiv \max_{\chi_0}|\mathcal{C}_i(\chi_0)|$. Figure~\ref{fig:res_strength} shows $\mathcal{C}_i^{\rm max}$ for the $3{:}1$, $5{:}2$, $2{:}1$, $3{:}2$, $4{:}3$ and $5{:}4$ resonances for a set of representative orbits spanning low to high eccentricity and inclination. For the configurations studied, the coefficients generally follow the approximate ordering:
\begin{equation}
3{:}2 > 2{:}1 > 5{:}2 > 3{:}1 \gg 5{:}4 \gg 4{:}3,
\label{eq:hierarchy}
\end{equation} 
with the exceptions discussed below, which agrees with previous studies~\cite{FHR,Berry}, although the specific values of the coefficients are different. 

The ordering in Eq.~\eqref{eq:hierarchy}, which refers to the fractional flux modifications, is consistent with two competing effects: symmetry-induced cancellations for odd $\beta_{r}$ and the degree of sampling of the underlying $2$-torus. As discussed in the previous section, odd $\beta_r$ resonances contain interference terms which necessarily cancel due to the discrete equatorial symmetry of Kerr, while, for even $\beta_r$, all terms contribute to the resonant flux modifications. Consequently, one expects resonances with $\beta_r=2$ to be stronger than those with $\beta_r=1$, and $\beta_r=4$ stronger than $\beta_r=3$. At the same time, resonances with smaller integer indices sample the $2$-torus to a lesser degree and are therefore expected to exhibit stronger resonant flux modifications~\cite{FlanaganHinderer2,Berry,SperiGair}. For a fixed $\beta_r$, increasing $\beta_\theta$ increases the number of polar oscillations before the trajectory closes on the radial–polar torus, allowing greater averaging along the resonant orbit.

Together, these two effects favor the $3{:}2$ as the strongest resonance: it combines the smallest even $\beta_r$ with the lowest allowed corresponding $\beta_\theta$. Although $5{:}2$ is the next resonance with $\beta_r=2$, the ordering $2{:}1>5{:}2$ suggests empirically that the lesser torus sampling of the $2{:}1$ resonance outweighs the cancellations associated with its odd value of $\beta_r$. The $5{:}2$ is followed by $3{:}1$, the next resonance with $\beta_r=1$. Both the $4{:}3$ and the $5{:}4$ resonances have larger integer indices and are therefore expected to be weaker because of greater averaging along their resonant orbits. The ordering $5{:}4>4{:}3$ reflects the symmetry cancellations favoring even $\beta_r$.

The ordering \eqref{eq:hierarchy}, however, admits exceptions, as illustrated in the lower-right panel of Fig.~\ref{fig:res_strength}, wherein the $2{:}1$ resonance is shown to have a larger resonant $Q$ coefficient than the $3{:}2$ resonance. This exception arises due to cancellations between the phase-dependent horizon and infinity contributions that suppress the $3{:}2$ resonant flux modification. We discuss this further in Sec.~\ref{sec:suppression}.

\begin{figure}[]
\includegraphics[width=\linewidth]{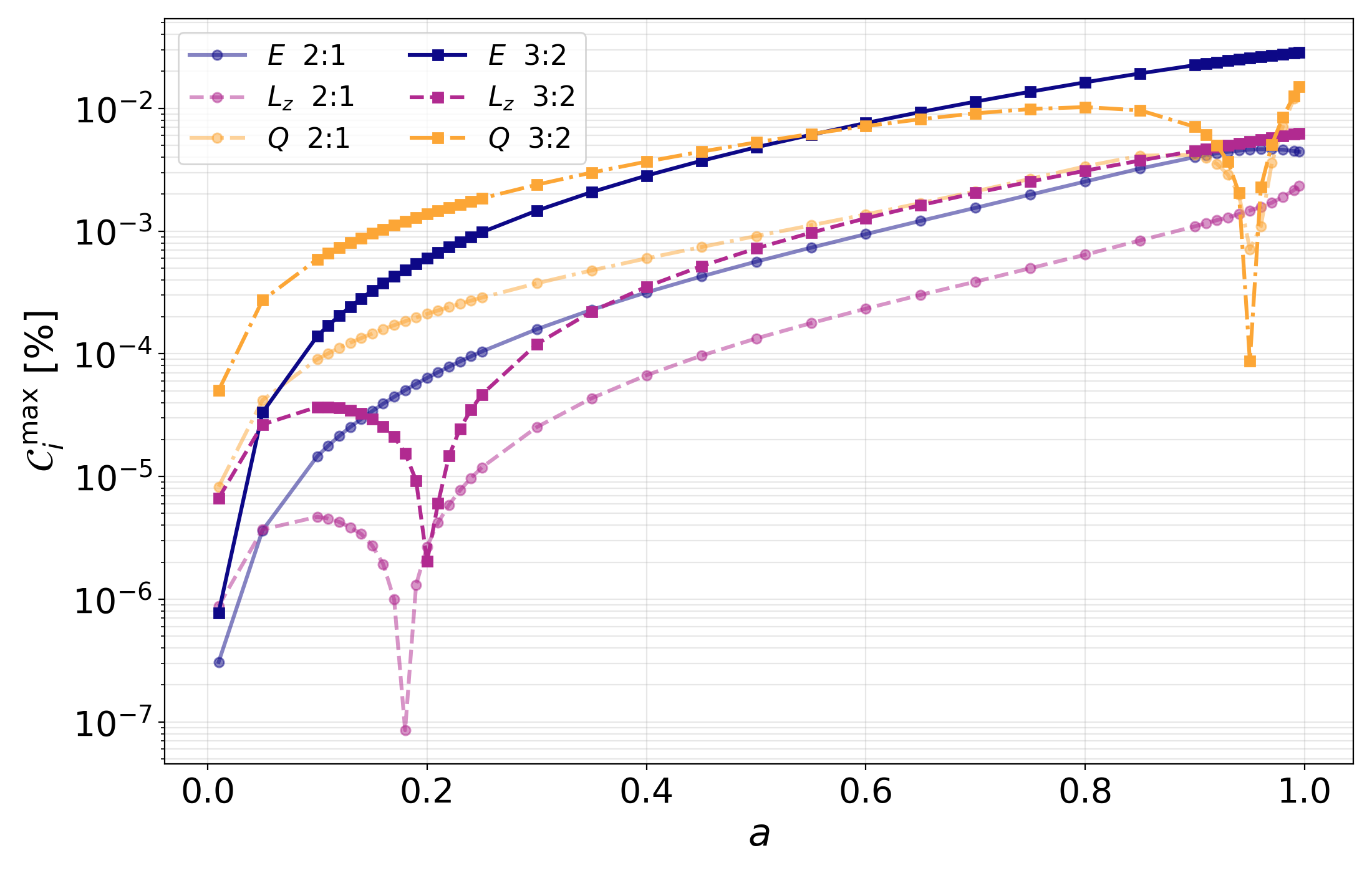}
\caption{Maximum resonance coefficients, $\mathcal{C}_i^{\rm max}\equiv\max_{\chi_0}|\mathcal{C}_i(\chi_0)|$, as functions of the primary BH spin $a$ for the $3{:}2$ (darker shading) and $2{:}1$ (lighter shading) resonances. We consider orbits with eccentricity $e=0.30$ and inclination $x_I=0.93969$, while the semi-latus rectum $p$ is adjusted to satisfy the resonance condition. Solid, dashed and dash-dotted lines correspond to the energy, axial angular momentum and Carter constant fluxes, respectively. All coefficients vanish in the Schwarzschild limit, $a\to0$. The low-spin minima in the $L_z$ coefficients and the high-spin minima in the $Q$ coefficients arise from cancellations between the signed resonant flux modifications at infinity and the horizon, as discussed in Sec.~\ref{sec:suppression}.}
\label{fig:spin_scaling}
\end{figure}

\subsection{Scaling with spin}

We now study the spin dependence of the resonant flux modifications for the $3{:}2$ and $2{:}1$ resonances. We vary the primary BH dimensionless spin $a\in(0,1)$ at fixed eccentricity $e=0.30$ and inclination $x_I=0.93969$, adjusting the semi-latus rectum $p$ to satisfy the resonance condition. 

Figure~\ref{fig:spin_scaling} shows the resulting maximum resonance coefficients, $\mathcal C_i^{\rm max}$. In the Schwarzschild limit $a\to0$, spherical symmetry removes the $q_{\theta}$ dependence of the forcing term $G_{i}^{(1)}$ in Eq.~\eqref{action} so the Fourier coefficients $G_{i,kn}(J)$ in Eq.~\eqref{eq:fourier_flux} vanish for $k\neq0$~\cite{FlanaganHinderer2}. Since the resonance condition $k\Omega_\theta - n\Omega_r=0$ requires both $k$ and $n$ to be nonzero, the corresponding harmonics with $n\neq0$ remain oscillatory and do not produce resonant flux modifications. 

At fixed eccentricity and inclination, increasing the spin requires a smaller semi-latus rectum $p$ to satisfy the resonance condition. The resulting resonant orbit is located more deeply in the strong-field, amplifying the modulus of the self-force Fourier coefficients $G_{i,kn}(J)$ and thereby the resonant flux modifications (on average).

As illustrated in Fig.~\ref{fig:spin_scaling}, however, the axial angular momentum and Carter constant resonance coefficients can exhibit local deviations from this average increase with spin. The $L_z$ coefficients exhibit local minima  around $a\simeq0.18$ and $a\simeq0.20$ for the $2{:}1$ and $3{:}2$ resonances, respectively, while the $Q$ coefficients exhibit local minima at $a\simeq0.95$ for both resonances. As discussed in Sec.~\ref{sec:suppression}, these minima occur due to cancellations between the signed resonant flux modifications at infinity and the horizon, suppressing the total resonant flux modification. Such cancellations can locally alter the relative strengths of the $3{:}2$ and $2{:}1$ resonances in parameter space. 

\begin{figure}[]
\centering
\includegraphics[width=\linewidth]{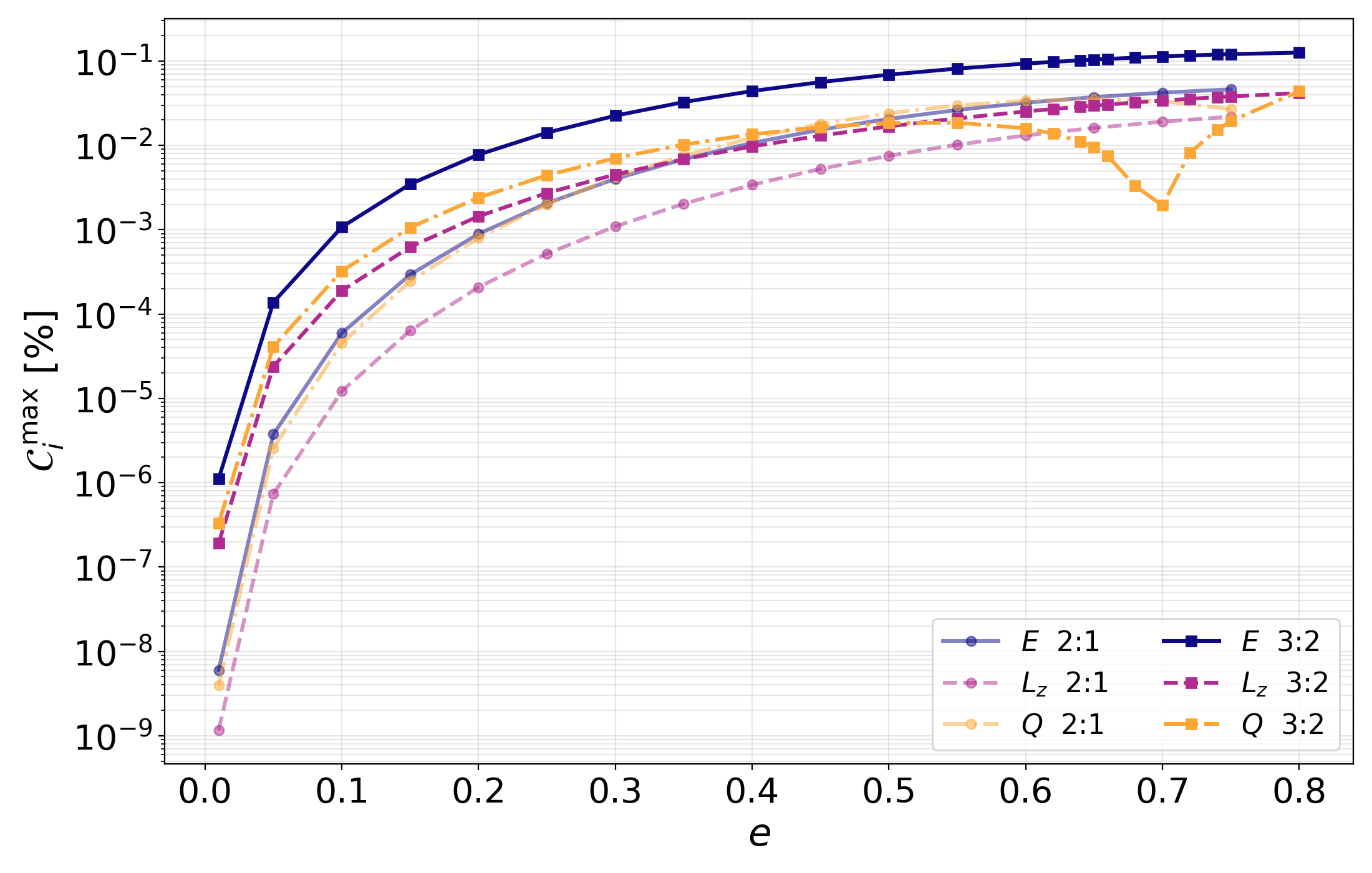}
\caption{Maximum resonance coefficients, $\mathcal{C}_i^{\rm max}\equiv\max_{\chi_0}|\mathcal{C}_i(\chi_0)|$, as functions of the eccentricity $e$ for the $3{:}2$ (darker shading) and $2{:}1$ (lighter shading) resonances. The primary BH spin and inclination are fixed at $a=0.90$ and $x_I=0.93969$, while the semi-latus rectum $p$ is adjusted to satisfy the resonance condition. Solid, dashed and dash-dotted lines correspond to the energy, axial angular momentum and Carter constant fluxes, respectively. The $N$-mode convergence criterion of Appendix~\ref{sec:appendix_B} is satisfied up to $e=0.75$ and $e=0.80$ for the $2{:}1$ and $3{:}2$ resonances, respectively. All coefficients vanish in the circular-orbit limit, $e\to0$. For the $3{:}2$ resonance, the local minimum in the $Q$ coefficient at $e\simeq0.7$ arises from cancellations between the signed resonant flux modifications at infinity and the horizon, as discussed in Sec.~\ref{sec:suppression}. As a result, over approximately $e\in[0.45,0.75]$, the $2{:}1$ resonance yields a larger $Q$ coefficient than the $3{:}2$ resonance.}
\label{fig:ecc_scaling}
\end{figure}

\subsection{Scaling with eccentricity}
\label{sec:scaling_eccentricity}

We next investigate the eccentricity dependence of the resonant flux modifications for the $3{:}2$ and $2{:}1$ resonances, varying $e$ at fixed primary BH spin $a=0.90$ and inclination $x_I=0.93969$.

At high eccentricities, the broader radial-harmonic spectrum requires more $n$-modes, while oscillatory source integrals become more expensive to evaluate accurately~\cite{Barton:2008,Fujita:2009,Chen:2026}. Recent work has begun to address these difficulties using specialized integration schemes designed to resolve highly oscillatory source integrands~\cite{Chen:2026}. 
An improved scheme for evaluating the source integrals in frequency-domain Teukolsky calculations was recently developed in Ref.~\cite{DHAROHighEccentricit}. Here, we restrict our analysis to eccentricities for which the $N$-mode convergence criterion of Appendix~\ref{sec:appendix_B} is satisfied: up to $e=0.75$ and $e=0.80$ for the $2{:}1$ and $3{:}2$ resonances, respectively. 

Figure~\ref{fig:ecc_scaling} shows the maximum resonance coefficients, $\mathcal C_i^{\rm max}$. The resonant flux modifications vanish as $e\to0$~\cite{FlanaganHinderer2,Drasco:2005kz}. In the quasi-circular limit, there is no radial oscillation and Teukolsky amplitudes with $n\ne0$ vanish, so the resonance condition $k\Omega_\theta - n\Omega_r=0$ cannot be non-trivially satisfied. As the orbits become more eccentric, the maximum resonance coefficients increase. Larger eccentricities broaden the spectrum of radial $n$-mode harmonics which contribute to the resonant fluxes, allowing for stronger resonant flux modifications. For the cases considered, increasing the eccentricity leads to small changes in the semi-latus rectum for the resonance condition to be satisfied, so that the broadening of the $n$-mode spectrum appears to drive the increase of the resonant modifications. For example, $p$ changes from $5.31M$ to $5.54M$ for the $3{:}2$ as $e$ is increased from $0.01$ to $0.8$.

The $3{:}2$ resonance coefficient for $Q$ exhibits a non-monotonic dependence on eccentricity, with a local minimum around $e\simeq0.7$. This suppression is consistent with cancellations between the phase-dependent infinity and horizon contributions, as illustrated in Sec.~\ref{sec:suppression}.  Consequently, over approximately $e\in[0.45,0.75]$, the $2{:}1$ resonance produces a larger $Q$ coefficient than the $3{:}2$ resonance.

\subsection{Scaling with orbital inclination}

\begin{figure}[]
\centering
\includegraphics[width=\linewidth]{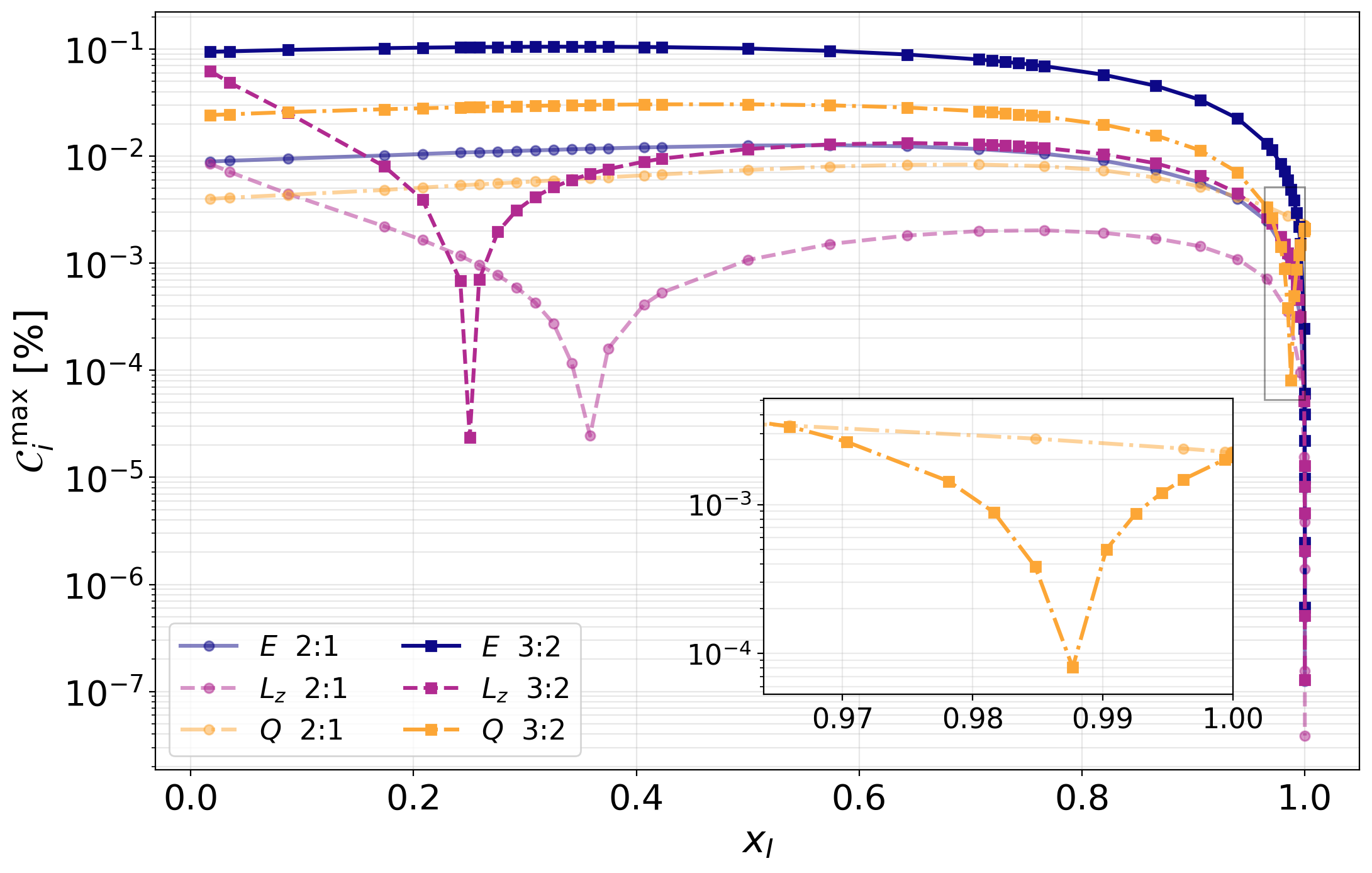}
\caption{Maximum resonance coefficients, $\mathcal{C}_i^{\rm max}\equiv\max_{\chi_0}|\mathcal{C}_i(\chi_0)|$, as functions of the inclination parameter $x_I$ for the $3{:}2$ (darker shading) and $2{:}1$ (lighter shading) resonances. We consider orbits with primary BH spin $a=0.90$ and eccentricity $e=0.30$, while the semi-latus rectum $p$ is adjusted to satisfy the resonance condition. Solid, dashed and dash-dotted lines correspond to the energy, axial angular momentum and Carter constant fluxes, respectively. The $E$ and $L_z$ coefficients vanish as $x_I\to1$. In the equatorial limit, the $Q$ coefficient approaches a constant value, since both the resonant contribution and the corresponding non-resonant flux entering its definition approach zero at the same rate. The minima in the $L_z$ coefficients at low $x_I$ and the minimum in the $Q$ coefficient for the $3{:}2$ resonance at $x_I\simeq0.987$ arise from cancellations between the signed resonant flux modifications at infinity and the horizon, as discussed in Sec.~\ref{sec:suppression}.}
\label{fig:theta_scaling}
\end{figure}

Lastly, we investigate the inclination dependence of the $3{:}2$ and $2{:}1$ resonances, varying $x_I\in(0,1)$ at fixed primary BH spin $a=0.90$ and eccentricity $e=0.30$. Figure~\ref{fig:theta_scaling} shows the resulting maximum resonance coefficients, $\mathcal C_i^{\rm max}$.

In the equatorial limit, $x_I\to1$, there is no polar oscillation and the Teukolsky amplitudes with $k\ne0$ vanish, prohibiting the resonance condition from being non-trivially satisfied~\cite{FlanaganHinderer2,Drasco:2005kz}. This suppression is reflected in the behavior of the $E$ and $L_z$ coefficients. As $x_I\to1$, however, the $Q$ coefficient approaches a constant value and requires separate limiting and numerical analysis, since both the resonant contribution and the corresponding non-resonant flux in the numerator and denominator of Eq.~\eqref{eq:coeffs} approach zero at the same rate.

As the orbit becomes more inclined, the spectrum of polar harmonics broadens, increasing the strength of the resonant modifications. As $x_{I}$ decreases from unity, the $E$ coefficient initially increases, reaches a maximum, and thereafter slightly decreases as the polar limit is approached. The maxima occur around $x_I\simeq0.57$ and $x_I\simeq0.34$ for the $2{:}1$ and $3{:}2$ resonances, respectively. The $Q$ coefficient exhibits similar behavior, with maxima around $x_I\simeq0.71$ and $x_I\simeq0.50$. At fixed spin and eccentricity, decreasing $x_I$ requires a larger semi-latus rectum $p$ to satisfy the resonance condition. The resulting outward shift into the weak-field thus appears to offset the broadening of the polar harmonic spectrum at large inclinations for the $E$ and $Q$ resonant flux modifications. For $L_z$, however, this does not occur as the resonance coefficient increases appreciably in the polar limit.

As observed in the scaling with spin and eccentricity, local minima occur in the resonance coefficients as the inclination is increased. The $L_{z}$ coefficient decreases to a minimum around $x_I\simeq0.36$ and $x_I\simeq0.25$ for the $2{:}1$ and $3{:}2$ resonances. The $3{:}2$ coefficient for $Q$ also shows a minimum near $x_I\simeq0.987$. These minima are consistent with cancellations between the phase-dependent infinity and horizon contributions, as discussed in the next subsection.

\begin{figure*}[]
\centering
\includegraphics[width=0.75\textwidth]{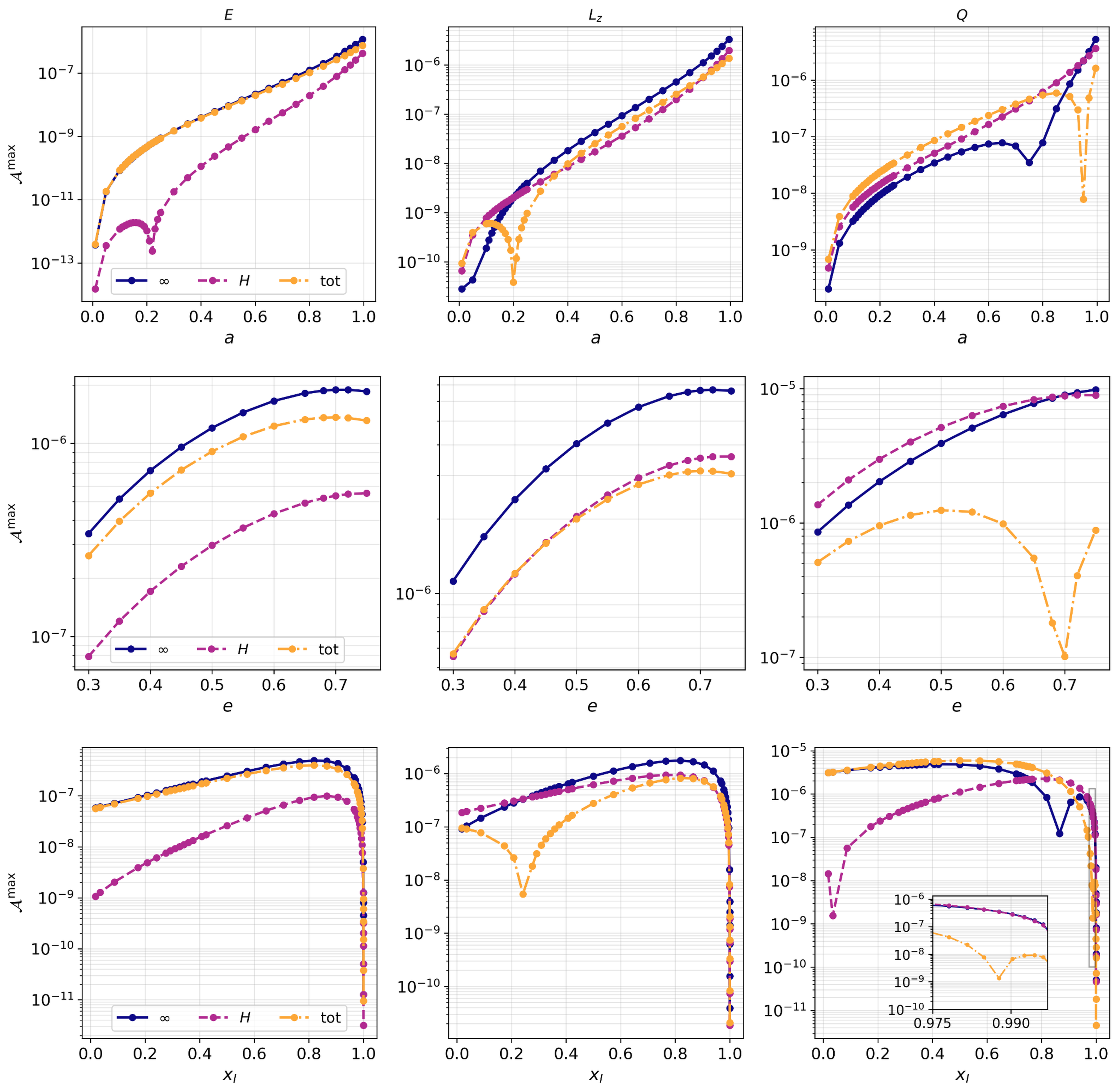}
\caption{Spin (upper row), eccentricity (middle row) and orbital inclination (lower row) dependence of the maximum resonant flux modifications for the $3{:}2$ resonance in Figs.~\ref{fig:spin_scaling},~\ref{fig:ecc_scaling} and~\ref{fig:theta_scaling}. From left to right, the columns correspond to the $E$, $L_z$ and $Q$ modifications. Solid, dashed and dash-dotted curves denote the infinity, horizon and total contributions, respectively. For each contribution, we plot the resonant flux amplitude $\mathcal{A}_i^{\sigma,\max}\equiv\max_{\chi_0} |\dot{J}_i^{\sigma}(\chi_0)-\langle\dot{J}_i^{\sigma}\rangle_{\chi_0}|$. The minima in the total $L_z$ and $Q$ modifications occur where the infinity and horizon contributions have comparable amplitudes and are out of phase. For the Carter constant, the crossing around $x_I\simeq0.75$ does not produce a minimum in the total modification because the infinity and horizon contributions are in phase. No cancellations are observed for $E$, for which the infinity contribution dominates the one at the horizon in the cases considered.}
\label{fig:resonant_flux_components}
\end{figure*}

\subsection{Suppression of resonant modifications}
\label{sec:suppression}

As discussed above, local minima occur in the orbital-parameter space for the resonant flux modifications to the angular momentum and Carter constant, as shown in Figs.~\ref{fig:spin_scaling}, \ref{fig:ecc_scaling}, and~\ref{fig:theta_scaling}. Such minima are not observed for $E$. Our calculations provide evidence that these minima, which, to the best of our knowledge, have not been previously reported, result from (at least partial) cancellations between the modifications to the flux to infinity and across the horizon.

We separate the phase-dependent flux modifications into their infinity and horizon contributions and define
\begin{equation}
\mathcal{A}_i^{\sigma}(\chi_0)
\equiv
\dot{J}_i^{\sigma}(\chi_0)
-\langle\dot{J}_i^{\sigma}\rangle_{\chi_0},
\label{eq:suppression_phase_deviations}
\end{equation}
where $i\in\{E,L_z,Q\}$ labels the orbital integral,
$\sigma\in\{\infty,H,\mathrm{tot}\}$ labels the flux channel, with
$\dot{J}_i^{\mathrm{tot}}=\dot{J}_i^\infty+\dot{J}_i^H$ and $\langle\dot{J}_i^\sigma\rangle_{\chi_0}$ denotes the non-resonant flux obtained by omitting interference terms between degenerate harmonics. These definitions imply $\mathcal{A}_i^{\mathrm{tot}}=\mathcal{A}_i^\infty+\mathcal{A}_i^H$ and we characterize each modification by its maximum over the phase, $\mathcal{A}_i^{\sigma,\max}\equiv
\max_{\chi_0}|\mathcal{A}_i^\sigma(\chi_0)|$.

Figure~\ref{fig:resonant_flux_components} shows these maxima as functions of the primary BH spin, eccentricity and inclination for the $3{:}2$ resonance. The minima in the total $L_z$ and $Q$ modifications are consistent with cancellations between infinity and horizon contributions which are of comparable amplitude but combine out of phase. Comparable amplitudes, however, do not always lead to cancellations: for the Carter constant, the amplitude crossing around $x_I\simeq0.75$ does not produce a minimum in the total modification since the individual contributions do not combine out of phase. That is, the relative phase dependence of the contributions also matters in addition to their respective amplitudes.

For the resonant modifications to the energy flux, we find for the orbits considered that the infinity contribution always dominates the horizon contribution, preventing any net suppression of the total modification. This horizon-infinity hierarchy is not always respected by the angular-momentum and Carter-constant modifications. Each flux channel depends on the same resonant amplitudes $Z^\star_{\ell mN}(\chi_0)$, but with different modal weights and an additional $Y^\star_{\ell mN}$-dependent term for $Q$. The materially different morphology of the energy-flux modifications compared to those of the angular momentum and Carter constant appears to arise from this non-trivial sum reweighting.

In particular, Eqs.~(\ref{eq:Einf_res}--\ref{eq:LzH_res}) give mode by mode
\begin{equation}
\dot L^\star_{z,\ell mN}(\chi_0)
=
\frac{m}{\omega_{mN}}
\dot E^\star_{\ell mN}(\chi_0),
\end{equation}
for $\omega_{mN}\ne0$. The resonant energy and angular momentum flux modifications therefore differ due to a $\chi_{0}$-independent reweighting of the same mode sum. The factor $m/\omega_{mN}$ removes the axisymmetric $m=0$ modes from $\dot L_z$, reweights the remaining modes, suppressing those with $|m|\ll N$, and is negative when $m\omega_{mN}<0$. The Carter-constant flux contains an additional source of phase dependence. Writing
$\mathcal{L}_{mN}
=m\langle\cot^2\theta\rangle L_z
-a^2\omega_{mN}\langle\cos^2\theta\rangle E$,
the term proportional to $|Z^\star_{\ell mN}|^2$ in
Eqs.~(\ref{eq:Qinf_res}--\ref{eq:QH_res}) can be written as
\begin{equation}
2\langle\cot^2\theta\rangle L_z\,
\dot L^\star_{z,\ell mN}
-
2a^2\langle\cos^2\theta\rangle E\,
\dot E^\star_{\ell mN}.
\end{equation}
The remaining contribution is
\begin{equation}
\frac{\Upsilon_\theta}{2\pi\omega_{mN}^3}
\operatorname{Re}
\left[
Z^\star_{\ell mN}(\chi_0)
\overline{Y^\star_{\ell mN}(\chi_0)}
\right],
\end{equation}
with an additional factor $\alpha_{\ell mN}$ for the horizon channel. As defined in
Eq.~\eqref{eq:resonant_Y}, $Y^\star_{\ell mN}$ weights each degenerate harmonic by its polar index $k$, adding further complexity to the summation of the resonant amplitudes $Z^\star_{\ell mN}(\chi_0)$.

These differences in modal weighting permit different morphologies of the horizon and infinity modifications for the angular momentum and Carter constant in comparison to the energy. The observed minima occur near crossovers in the relative sizes of the two contributions, provided they combine out of phase. The crossover locations need not coincide for $L_z$ and $Q$, as illustrated by the spin scan in Fig.~\ref{fig:resonant_flux_components}.

\section{Conclusions}
\label{sec:conclusions}

In this work, we present, to our knowledge, the most extensive systematic analysis to date of resonant modifications to the energy, axial angular momentum, and Carter constant fluxes across a broad region of orbital parameter space. We use the publicly available \texttt{pybhpt} code~\cite{Nasipak:2022,Nasipak:2025tby,Nasipak_pybhpt_2026} to solve the frequency-domain Teukolsky equation and implement the resonant-flux formalism of Ref.~\cite{FHR}, coherently combining the degenerate harmonics at resonance.

We analytically show that the minimum number of cycles in the resonance coefficients over $\chi_0\in[0,2\pi)$ for a $\beta_\theta{:}\beta_r$ resonance is given by $\mathrm{lcm}(2,\beta_r)$. Beyond this selection rule, we identify three novel behaviors of the resonance coefficients. First, all resonances considered admit phases at which the $E$, $L_z$, and $Q$ coefficients vanish simultaneously. Second, the coefficients span all possible sign combinations across the configurations explored. Depending on the orbital parameters and resonance, all three fluxes can be enhanced, all three suppressed, or some enhanced and others suppressed relative to their non-resonant values. Third, owing to our systematic exploration of the parameter space, we found that the coefficients associated with $L_z$ and $Q$ can exhibit non-monotonic behavior, with local minima arising due to cancellations between the horizon and infinity contributions. Distinguishing total cancellation, resulting in vanishing resonance coefficients, from partial cancellation is challenging numerically. An interesting avenue for future work would be to explore whether the $L_{z}$ and $Q$ coefficients vanish exactly in these scenarios.

We also provide the first Teukolsky-based calculations of the $E$, $L_z$, and $Q$ resonance coefficients for the $5{:}2$ and $5{:}4$ resonances. Comparing the maximum coefficients over $\chi_0$, we find the approximate ordering
\begin{equation*}
3{:}2 > 2{:}1 > 5{:}2 > 3{:}1
\gg 5{:}4 \gg 4{:}3
\end{equation*}
across the configurations explored, which agrees with previous studies~\cite{FHR,Berry}. This ordering is consistent with a balance between symmetry-induced cancellations in the fluxes for odd $\beta_r$, and increased averaging along the resonant orbit as $\beta_\theta$ increases at fixed $\beta_r$. 

We also investigated the dependence of the $3{:}2$ and $2{:}1$ resonances on the primary BH spin, orbital eccentricity, and inclination. The resonant flux modifications vanish in the Schwarzschild, circular, and equatorial limits. Increasing the spin breaks spherical symmetry, allowing polar harmonics to contribute to phase-dependent resonant interference. Higher eccentricity broadens the radial-harmonic spectrum, while inclination broadens the polar-harmonic spectrum. These effects allow for stronger resonant modifications to the fluxes. Locally, however, the coefficients can vary non-monotonically: local minima arise in the $L_z$ and $Q$ coefficients due to destructive interference between the phase-dependent horizon and infinity contributions. These cancellations can locally reverse the $3{:}2>2{:}1$ ordering for the $L_{z}$ and $Q$ coefficients.

The astrophysical impact of a resonance crossing, particularly the accumulated GW dephasing, also depends on its location and duration and on the observation time~\cite{Levati:2026emw,Brink:2013nna,Brink:2015roa,Ruangsri:2013hra}. Our fluxes are infinite-time averages for a point particle on a resonant geodesic with fixed orbital parameters and a specified radial-polar phase. As noted in Ref.~\cite{FHR}, these quantities alone do not determine continuous evolution through resonance. The discontinuity between resonant and non-resonant geodesic fluxes is an artifact of infinite-time averaging, whereas the fluxes during a physical inspiral vary smoothly as the orbital frequencies and resonant phase evolve.

The phase dependence found here follows from the equatorial reflection symmetry of Kerr, Eq.~\eqref{eq:equatorial_symmetry}, motivating extensions to backgrounds without this symmetry~\cite{Cardoso:2018ptl,Chen:2022,Fransen:2022}. A recent study of Kerr symmetry breaking in EMRIs is presented in Ref.~\cite{Muguruza:2026hqn}. Generalized Teukolsky formalisms~\cite{Li:2022pcy,Cano:2023tmv} may enable calculations of low-order resonances in these spacetimes, testing whether the phase dependence persists and how the strength of the resonance flux modifications changes.

Together with our analytical results, the calculated coefficients provide input for calibrating localized, smooth, and phase-coherent corrections to adiabatic fluxes in phenomenological descriptions of resonance dynamics~\cite{SperiGair,Levati:2026emw}. Such models would support studies of resonance-induced parameter-estimation biases across a broad region of EMRI parameter space. A further extension is to model inspirals encountering several transient resonances while consistently evolving the radial-polar phase between crossings. Our numerical framework is also well suited to constructing datasets of orbit- and phase-dependent flux modifications for future LISA waveform models and data-analysis studies.

\begin{acknowledgments}
We are especially grateful to Scott Hughes for his help throughout this work, including numerous discussions and email exchanges, detailed feedback and clarifications, and data that enabled us to validate several of our calculations. We thank Maarten van de Meent, Zachary Nasipak and Niels Warburton for useful comments and discussions. L.K. and A.C.-A. thank Dorothy Cutler for insightful discussions. E.L., L.K. and A.C.-A. acknowledge support from NASA through the award 80NSSC26K1237. ChatGPT v$5.6$ by OpenAI and Claude Fable $5$ by Anthropic were used in this work to accelerate numerical implementation and proofreading. All scientific analyses, numerical results and conclusions were produced by the authors, who take full responsibility for the contents of this work. Computations were performed using the Wake Forest University (WFU) High Performance Computing Facility, a centrally managed computational resource available to WFU researchers, including faculty, staff, students and collaborators~\cite{WakeHPC}. 
\end{acknowledgments}

\appendix

\section{Phase Dependence of the Resonant Flux Modifications}
\label{sec:appendix_A}

In this Appendix, we derive the following selection rule: the minimum number of cycles of the resonant flux modifications over $\chi_{0}\in[0,2\pi)$ is given by $\mathrm{lcm}(2,\beta_{r})$, where $\mathrm{lcm}$ denotes the least common multiple. 

In the following, we consider the energy flux to future null infinity. We expect our conclusions to hold for the energy down-horizon flux as well as for the axial angular momentum and Carter constant fluxes. Substituting Eq.~(\ref{eq:resonant_Z_j_squared}) into Eq.~(\ref{eq:Einf_res}) and expanding the squared modulus, we obtain
\begin{widetext}
\begin{align}
4\pi\omega_{mN}^2\dot{E}_{\ell m N}^{\infty}(\chi_{0}) &=
\ab{\,\sum_{j} e^{i\xi_{mk_jn_j}(\chi_0)}\,\check{Z}^{\rm in}_{\ell m k_j n_j}}^2=
\sum_{j,j'} e^{i\br{\xi_{mk_jn_j}(\chi_0)-\xi_{mk_{j'}n_{j'}}(\chi_0)}}
\check{Z}^{\rm in}_{\ell m k_j n_j}\bcZk{\ell m k_{j'} n_{j'}}\nonumber\\
&=\sum_{j}
\ab{\cZk{\ell m k_j n_j}}^2+\sum_{j\neq j'} e^{i\br{\xi_{mk_jn_j}(\chi_0)-\xi_{mk_{j'}n_{j'}}(\chi_0)}}
\cZk{\ell m k_j n_j}\bcZk{\ell m k_{j'} n_{j'}},\label{eq:ElmN}
\end{align}
\end{widetext}
which is equivalent to Eq.~(4.2) of Ref.~\cite{FHR}. Using Eq.~(\ref{eq:res_family}) and Eq.~(\ref{eq:delta_phase_factor}), we rewrite Eq.~(\ref{eq:ElmN}) as 
\begin{align}
    4\pi\omega_{mN}^2\dot{E}_{\ell m N}^{\infty}(\chi_{0}) &= \dot{E}_{\ell m N}^{\infty,\text{no-res}}+\dot{E}_{\ell m N}^{\infty,\rm{res}}(\chi_{0}),
\end{align}
where
\begin{align}
\dot{E}_{\ell m N}^{\infty,\text{no-res}}&=\sum_{j} \ab{\cZk{\ell m k_j n_j}}^2,\\ 
\dot{E}_{\ell m N}^{\infty,\rm{res}}(\chi_{0})&=\sum_{j\neq j'} e^{is(j,j') q_{\theta0}(\chi_0)}\,
\cZk{\ell m k_j n_j}\bcZk{\ell m k_{j'} n_{j'}} \label{eq:interference}, 
\end{align}
and we have defined
\begin{equation}
s(j, j') = (j - j')\,\beta_r \in \cu{\pm\beta_r, \pm 2\beta_r, \pm 3\beta_r, \dots}.
\label{eq:sdef}
\end{equation}

Under the discrete equatorial symmetry $\theta \to \pi - \theta$ of the Kerr spacetime, we have from Eq.~$(3.52)$ of Ref.~\cite{Drasco:2005kz}
\begin{equation}
\check{Z}^{\star}_{\ell,-m,-k,-n}
= (-1)^{\ell+k}\bar{\check{Z}}^{\star}_{\ell m k n}.
\label{eq:kerrsym}
\end{equation}
The resonant frequencies satisfy
\begin{equation}
\omega_{-m,-N} = -\omega_{mN}, \qquad \omega_{-m,-N}^2 = \omega_{mN}^2.
\end{equation}
This symmetry can be used to reduce the sum over $(m,N)$ in Eq.~(\ref{eq:Einf_res}) of the resonant contribution to a sum over $(m>0,N\in\mathbb{Z})$ and $(m=0,N>0)$. In particular,  under $N \to -N$, $k_{0}$ and $n_{0}$ [defined in Eq.~(\ref{eq:res_family})] can be taken to obey
\begin{equation}
(k_0, n_0) \to (-k_0, -n_0).
\end{equation}
Under $(m,N) \to (-m,-N)$, we thus have
\begin{equation}
k_{\bar{j}} = -k_0 + \bar{j}\beta_r, \qquad
n_{\bar{j}} = -n_0 - \bar{j}\beta_\theta
\label{eq:barredmodes}
\end{equation}
and
\begin{align}
\dot{E}^{\infty,\rm{res}}_{\ell,-m,-N}(\chi_0)
&= \sum_{\bar{j}\neq\bar{j}'}
e^{is(\bar{j},\bar{j}')q_{\theta0}(\chi_0)}
\cZk{\ell,-m,k_{\bar{j}},n_{\bar{j}}}\nonumber\\
&\qquad\times\bcZk{\ell,-m,k_{\bar{j}'},n_{\bar{j}'}}.
\label{eq:negint}
\end{align}
Taking $\bar{j} \to -j$ and $\bar{j}' \to -j'$ in the sums,
\begin{align}
\dot{E}^{\infty,\rm{res}}_{\ell,-m,-N}(\chi_0)
&=\sum_{j\neq j'}
e^{-is(j,j')q_{\theta0}(\chi_0)}
\cZk{\ell,-m,-k_j,-n_j}\nonumber\\
&\qquad\times
\bcZk{\ell,-m,-k_{j'},-n_{j'}}.
\label{eq:negintrelabel}
\end{align}
Applying the symmetry~\eqref{eq:kerrsym} to both amplitudes,
\begin{widetext}
\begin{align}
\dot{E}^{\infty,\rm{res}}_{\ell,-m,-N}(\chi_0)
&=  \sum_{j\neq j'}
e^{-is(j,j')q_{\theta0}(\chi_0)}\,(-1)^{\ell+k_j}(-1)^{\ell+k_{j'}}\,
\bcZk{\ell m k_j n_j}\cZk{\ell m k_{j'} n_{j'}} \\
&=  \sum_{j\neq j'}
e^{-is(j,j')q_{\theta0}(\chi_0)}\,(-1)^{s(j,j')}\,
{\bcZk{\ell m k_j n_j}}\,\cZk{\ell m k_{j'} n_{j'}}.
\end{align}
Therefore
\begin{align}
\dot{E}^{\infty,\rm{res}}_{\ell m N}(\chi_0)
+ \dot{E}^{\infty,\rm{res}}_{\ell,-m,-N}(\chi_0)
&= 
\sum_{j\neq j'}
\br{e^{is(j,j')q_{\theta0}(\chi_0)}\,\cZk{\ell m k_j n_j}\,{\bcZk{\ell m k_{j'} n_{j'}}}
+ (-1)^{s(j,j')} e^{-is(j,j')q_{\theta0}(\chi_0)}\,{\bcZk{\ell m k_j n_j}}\,\cZk{\ell m k_{j'} n_{j'}}}.
\label{eq:pairsum}
\end{align}
The sum in Eq.~\eqref{eq:pairsum} is over ordered pairs. Writing the sum over unordered pairs with $j>j'$, we have 
\begin{align}
\dot{E}^{\infty,\rm{res}}_{\ell m N}(\chi_0)
+ \dot{E}^{\infty,\rm{res}}_{\ell,-m,-N}(\chi_0)
&= 
\sum_{j>j'}
\br{e^{is(j,j')q_{\theta0}(\chi_0)}\,\cZk{\ell m k_j n_j}\,{\bcZk{\ell m k_{j'} n_{j'}}}
+ (-1)^{s(j,j')} e^{-is(j,j')q_{\theta0}(\chi_0)}\,{\bcZk{\ell m k_j n_j}}\,\cZk{\ell m k_{j'} n_{j'}}}\nonumber\\
&\qquad\qquad +\pa{j\leftrightarrow j'}\nonumber\\
&=\sum_{j>j'}
2\br{1+\pa{-1}^{s(j,j')}}\mathrm{Re}\pa{e^{is(j,j')q_{\theta0}(\chi_0)}\,\cZk{\ell m k_j n_j}\,{\bcZk{\ell m k_{j'} n_{j'}}}}.
\label{eq:pairsum2}
\end{align}
Hence, only even $s(j,j')$ terms contribute to the sum in Eq.~\eqref{eq:pairsum2}, allowing us to write
\begin{align}
\dot{E}^{\infty,\rm{res}}_{\ell m N}(\chi_0)
+ \dot{E}^{\infty,\rm{res}}_{\ell,-m,-N}(\chi_0)
=
\sum_{j> j'}
4\,\mathrm{Re}\pa{e^{is(j,j')q_{\theta0}(\chi_0)}\,\cZk{\ell m k_j n_j}\,{\bcZk{\ell m k_{j'} n_{j'}}}},
\label{eq:evensum}
\end{align}
\end{widetext}
where the sum is taken over even $s(j,j')$. The periodicity (with respect to $q_{\theta0}$) of the sum of terms on the left-hand side of Eq.~\eqref{eq:evensum} is set by the smallest value of $\ab{s(j,j')}$. From Eq.~\eqref{eq:sdef}, we see that the minimum number of cycles over $q_{\theta0}\in[0,2\pi)$ is given by $\mathrm{lcm}(2,\beta_r)$. Since $q_{\theta0}$ and $\chi_{0}$ are different parametrizations of the same geometric circle, and $q_{\theta0}(\chi_{0})$ is a monotonic function of $\chi_{0}$, the number of cycles exhibited by the left-hand side of Eq.~\eqref{eq:evensum} with respect to $\chi_{0}$ and $q_{\theta0}$ over the interval $[0,2\pi)$ is the same. This phase dependence is independent of the mode $(\ell, m, N)$ and is thus inherited by the total flux \eqref{eq:Einf_res}.

\section{Numerical Convergence and Validation}
\label{sec:appendix_B}

In this Appendix, we describe the scheme we use to determine the truncation of the various infinite mode sums involved in the flux computations and to assess the numerical errors in the resulting resonant fluxes.

\subsection{Convergence of the resonant mode sums}
\label{sec:convergence}

Evaluating Eq.~\eqref{eq:total_res_flux} requires truncating both the coherent sum over degenerate modes within each resonant family and the outer mode sums. For a given $(\ell,m,N)$ family, the degenerate modes are parametrized by the integer $j$, with $-j_{\rm max}\leq j\leq j_{\rm max}$ and the outer sums are truncated at $-N_{\max}\leq N\leq N_{\rm max}$ and $2\leq\ell\leq\ell_{\rm max}$. For each $\ell$, $m$ is summed over the full range $-\ell \leq m\leq \ell$.

We determine $j_{\rm max}$, $N_{\rm max}$ and $\ell_{\rm max}$ adaptively by requiring corrections to the resonant ($E,L_z,Q$) fluxes below some tolerance when incrementing the mode bounds. We refer to such an increment of one of $(j_{\rm max},N_{\rm max},\ell_{\rm max})$ as a shell. For a shell $S$ added along one of the three truncation directions, we define the shell residual $r_{S}$ as
\begin{equation}
r_{S}
=
\frac{
\displaystyle
\max_{\chi_0}
\left|
\Delta \dot J_S^{\sigma}(\chi_0)
\right|
}{
\displaystyle
\max_{\chi_0}
\left|
\dot J^{\sigma}(\chi_0)
\right|
},
\end{equation}
where $\sigma\in\{\infty,H,\mathrm{tot}\}$, $J\in \{E,L_z,Q\}$, $\Delta\dot J_S^{\sigma}$ is the contribution of the added shell and $\dot J^{\sigma}$ is the flux including that shell. The maxima in the numerator and denominator are taken over the full range of sampled values of $\chi_0$. Our criterion for a converged shell is when $r_{S} < 10^{-5}.$

Shell convergence is evaluated first in $j_{\rm max}$, then in $N_{\rm max}$, and lastly in $\ell_{\rm max}$, where the three truncations are not treated independently. At a given candidate value of $\ell_{\max}$ and $N_{\max}$, we first determine $j_{\rm max}$. Then, convergence of an added shell to $N_{\max}$ is tested. Upon failure, $N_{\max}$ is incremented and $j_{\max}$ updated, which we repeat until convergence in $N$ has been reached. Shell convergence in $\ell$ is similarly determined, incrementing $N_{\max}$ and, in turn, $j_{\max}$ until the convergence criterion has been met for $\ell$. The procedure ends when the convergence criterion in $\ell$ is satisfied.

We require the convergence criterion to be satisfied by two consecutive complete $j$ shells, five consecutive complete $N$ shells, and two consecutive complete $\ell$ shells.

\subsection{Error estimation of the resonant fluxes}

Once the truncation procedure described in the previous section has been satisfied, we subsequently assess the numerical errors in the fluxes. We first compute the total resonant fluxes using \texttt{pybhpt}'s default geodesic resolution $n_{\rm samples}^{\rm geo}=1024$. We then assess the validity of the calculation by monitoring the relative error of the individual Teukolsky amplitudes entering each $(\ell,m,N)$ family.

For each complex amplitude $Z^\star_{\ell mkn}$, we use \texttt{pybhpt}'s estimate of its relative error~\cite{Nasipak_pybhpt_2026}
\begin{equation}
\epsilon^\star_{\ell mkn}
=
\begin{cases}
\epsilon^\star_{\rm conv},
& \epsilon^\star_{\rm conv}\leq10^{-2},\\[4pt]
\max\!\left(\epsilon^\star_{\rm conv},\epsilon^\star_{\rm canc}\right),
& \epsilon^\star_{\rm conv}>10^{-2},
\end{cases}
\end{equation}
where $\epsilon^\star_{\rm conv}$ estimates the integration error from grid refinement and $\epsilon^\star_{\rm canc}$ accounts for potential catastrophic cancellation.

We flag an amplitude as potentially problematic when $\epsilon^\star_{\ell mkn}>10^{-3}$. All computed $Z^\star_{\ell mkn}$ are retained when constructing the resonant families and determining $j_{\rm max}$, $N_{\rm max}$ and $\ell_{\rm max}$. After the truncated resonant mode sum has met the convergence criteria, we assess whether the flagged amplitudes are significant by computing
\begin{equation}
\delta
=
\frac{
\displaystyle
\max_{\chi_0}
\left|
\Delta\dot J^\sigma(\chi_0)
\right|
}{
\displaystyle
\max_{\chi_0}
\left|
\dot J^{\sigma}(\chi_0)
\right|
},
\end{equation}
where $\Delta\dot J^\sigma$ is the contribution from either a single flagged amplitude or all flagged amplitudes and $\dot J^{\sigma}$ is the total flux including all flagged and non-flagged amplitudes. We compute $\delta$ separately for each flagged amplitude and for all flagged amplitudes combined in order to assess their individual and collective effect on the fluxes.

We impose the criterion $\delta < 10^{-5}$ on the fluxes. If this is violated, the calculation is repeated at progressively higher geodesic resolutions, $n_{\rm samples}^{\rm geo} = (2048, 4096, \ldots)$, and the  $\delta$-assessment is repeated for each. If convergence in $j_{\rm max}$, $N_{\rm max}$, or $\ell_{\rm max}$ is not achieved, or if significant amplitude flags remain after reaching a chosen maximum geodesic resolution, the computation stops. In practice, for the orbits considered in this work, the maximum geodesic resolution required has $n_{\rm samples}^{\rm geo}=2048$.

\subsection{Validation}

We compute the resonant flux modifications for the resonances and orbital configurations considered in Ref.~\cite{FHR}. For each configuration, we list the values of $\ell_{\rm max}$, $N_{\rm max}$, and $j_{\rm max}$ in Table~\ref{tab:convergence_parameters}. We report in Tables~\ref{tab:fhr_comparison_1}--\ref{tab:fhr_comparison_4} the peak-to-trough variation~\cite{FHR}
\begin{equation}
  \Delta_i^{\sigma} \equiv \left| \frac{\bigl|\dot{J}^{\sigma}_{i,\max}\bigr| - \bigl|\dot{J}^{\sigma}_{i,\min}\bigr|}{\left(\bigl|\dot{J}^{\sigma}_{i,\max}\bigr| + \bigl|\dot{J}^{\sigma}_{i,\min}\bigr|\right)/2} \right|,
  \label{eq:DeltaFHR}
\end{equation}
where $i\in\{E,L_z,Q\}$ labels the orbital integral and $\sigma\in\{\infty,H,\mathrm{tot}\}$ labels the flux channel. For a selected subset of orbits, we also compared our results with new calculations performed using an updated version of \texttt{gremlin}, the code used to compute the Teukolsky amplitudes in Ref.~\cite{FHR}. For the $3{:}2$ and $2{:}1$ resonances, we find excellent agreement, with relative differences in the resonant fluxes on the order of $10^{-5}$. Across the selected cases, the maximum relative differences in the resonant fluxes are of the order $10^{-3}$, $10^{-2}$, and $10^{-2}$ for $E$, $L_z$, and $Q$, respectively.

\begin{table*}[]
\centering
\caption{Truncated mode-sum parameters $(\ell_{\rm max},N_{\rm max},j_{\rm max})$ for the resonances and orbital configurations of Tables~\ref{tab:fhr_comparison_1}--\ref{tab:fhr_comparison_4}, with primary BH spin $a=0.90$. Details about the convergence criteria can be found in Appendix~\ref{sec:appendix_B}.}
\label{tab:convergence_parameters}
\begin{tabular}{ccccc}
\hline\hline
Resonance &
\begin{tabular}{c}
$e=0.3$\\
$x_I=0.34202$
\end{tabular} &
\begin{tabular}{c}
$e=0.3$\\
$x_I=0.93969$
\end{tabular} &
\begin{tabular}{c}
$e=0.7$\\
$x_I=0.34202$
\end{tabular} &
\begin{tabular}{c}
$e=0.7$\\
$x_I=0.93969$
\end{tabular} \\
\hline
$3{:}1$ & $(15,47,14)$ & $(21,35,10)$ & $(17,121,19)$ & $(24,129,10)$ \\
$2{:}1$ & $(13,32,12)$ & $(17,26,10)$ & $(15,99,17)$  & $(21,106,10)$ \\
$3{:}2$ & $(11,42,10)$ & $(13,32,10)$ & $(12,142,10)$ & $(15,151,10)$ \\
$4{:}3$ & $(10,53,10)$ & $(12,42,10)$ & $(11,173,10)$ & $(13,188,10)$ \\
\hline\hline
\end{tabular}
\end{table*}

\begin{table*}
\centering
\caption{Peak-to-trough variations of the resonant fluxes for the orbits considered in Ref.~\cite{FHR}, with primary BH spin $a=0.90$, eccentricity $e=0.30$ and inclination $x_I=0.34202$, corresponding to $\theta_{\min}=20^\circ$ in the convention of Ref.~\cite{FHR}. For the energy, axial angular momentum and Carter constant fluxes, we report $\Delta_i^{\sigma}$ as defined in Eq.~\eqref{eq:DeltaFHR}, where $i\in\{E,L_z,Q\}$ labels the orbital integral and $\sigma\in\{\infty,H,\mathrm{tot}\}$ labels the flux channel.}
\label{tab:fhr_comparison_1}
\renewcommand{\arraystretch}{1.12}
\setlength{\tabcolsep}{3.2pt}
\resizebox{\textwidth}{!}{%
\begin{tabular}{
c c c c
c c c
c c c
c c c
}
\toprule
&
&
&
&
\multicolumn{3}{c}{Horizon $[\%]$}
& \multicolumn{3}{c}{Infinity $[\%]$}
& \multicolumn{3}{c}{Total $[\%]$}
\\
\cmidrule(lr){5-7}
\cmidrule(lr){8-10}
\cmidrule(lr){11-13}
Res.
& $e$
& $x_I$
& $p/M$
& $\Delta_{E}^H$
& $\Delta_{L_z}^H$
& $\Delta_{Q}^H$
& $\Delta_{E}^{\infty}$
& $\Delta_{L_z}^{\infty}$
& $\Delta_{Q}^{\infty}$
& $\Delta_{E}^{\rm tot}$
& $\Delta_{L_z}^{\rm tot}$
& $\Delta_{Q}^{\rm tot}$
\\
\midrule
$3{:}1$
& 0.3
& 0.34202
& 5.04884
& $5.3\times10^{-2}$
& $1.6\times10^{-2}$
& $1.1\times10^{-2}$
& $1.4\times10^{-3}$
& $8.9\times10^{-4}$
& $1.1\times10^{-3}$
& $1.8\times10^{-3}$
& $2.9\times10^{-4}$
& $1.2\times10^{-3}$
\\
$2{:}1$
& 0.3
& 0.34202
& 6.12789
& $7.4\times10^{-1}$
& $2.6\times10^{-1}$
& $3.1\times10^{-1}$
& $2.0\times10^{-2}$
& $9.9\times10^{-3}$
& $1.1\times10^{-2}$
& $2.3\times10^{-2}$
& $2.3\times10^{-4}$
& $1.2\times10^{-2}$
\\
$3{:}2$
& 0.3
& 0.34202
& 8.65334
& $1.0\times10^{1}$
& $5.1\times10^{0}$
& $6.9\times10^{0}$
& $2.3\times10^{-1}$
& $8.0\times10^{-2}$
& $5.4\times10^{-2}$
& $2.1\times10^{-1}$
& $1.2\times10^{-2}$
& $6.0\times10^{-2}$
\\
$4{:}3$
& 0.3
& 0.34202
& 11.31578
& $3.2\times10^{-7}$
& $2.6\times10^{-8}$
& $5.8\times10^{-7}$
& $5.1\times10^{-9}$
& $7.1\times10^{-10}$
& $2.5\times10^{-9}$
& $4.9\times10^{-9}$
& $8.6\times10^{-10}$
& $2.4\times10^{-9}$
\\
\bottomrule
\end{tabular}%
}
\end{table*}

\begin{table*}[]
\centering
\caption{Peak-to-trough variations of the resonant fluxes for the orbits considered in Ref.~\cite{FHR}, with primary BH spin $a=0.90$, eccentricity $e=0.30$ and inclination $x_I=0.93969$, corresponding to $\theta_{\min}=70^\circ$ in the convention of Ref.~\cite{FHR}. For the energy, axial angular momentum and Carter constant fluxes, we report $\Delta_i^{\sigma}$ as defined in Eq.~\eqref{eq:DeltaFHR}, where $i\in\{E,L_z,Q\}$ labels the orbital integral and $\sigma\in\{\infty,H,\mathrm{tot}\}$ labels the flux channel.}
\label{tab:fhr_comparison_2}
\renewcommand{\arraystretch}{1.12}
\setlength{\tabcolsep}{3.2pt}
\resizebox{\textwidth}{!}{%
\begin{tabular}{
c c c c
c c c
c c c
c c c
}
\toprule
&
&
&
&
\multicolumn{3}{c}{Horizon $[\%]$}
& \multicolumn{3}{c}{Infinity $[\%]$}
& \multicolumn{3}{c}{Total $[\%]$}
\\
\cmidrule(lr){5-7}
\cmidrule(lr){8-10}
\cmidrule(lr){11-13}
Res.
& $e$
& $x_I$
& $p/M$
& $\Delta_{E}^H$
& $\Delta_{L_z}^H$
& $\Delta_{Q}^H$
& $\Delta_{E}^{\infty}$
& $\Delta_{L_z}^{\infty}$
& $\Delta_{Q}^{\infty}$
& $\Delta_{E}^{\rm tot}$
& $\Delta_{L_z}^{\rm tot}$
& $\Delta_{Q}^{\rm tot}$
\\
\midrule
$3{:}1$
& 0.3
& 0.93969
& 2.91117
& $3.7\times10^{-3}$
& $7.6\times10^{-3}$
& $3.4\times10^{-2}$
& $1.1\times10^{-3}$
& $6.2\times10^{-4}$
& $2.2\times10^{-4}$
& $1.1\times10^{-3}$
& $3.5\times10^{-4}$
& $2.3\times10^{-3}$
\\
$2{:}1$
& 0.3
& 0.93969
& 3.55601
& $4.1\times10^{-2}$
& $7.5\times10^{-2}$
& $5.6\times10^{-1}$
& $8.8\times10^{-3}$
& $4.1\times10^{-3}$
& $4.9\times10^{-3}$
& $8.0\times10^{-3}$
& $2.2\times10^{-3}$
& $8.4\times10^{-3}$
\\
$3{:}2$
& 0.3
& 0.93969
& 5.34138
& $1.2\times10^{0}$
& $7.5\times10^{-1}$
& $1.4\times10^{1}$
& $5.8\times10^{-2}$
& $1.8\times10^{-2}$
& $2.4\times10^{-2}$
& $4.5\times10^{-2}$
& $9.0\times10^{-3}$
& $1.4\times10^{-2}$
\\
$4{:}3$
& 0.3
& 0.93969
& 7.41979
& $1.1\times10^{-8}$
& $1.8\times10^{-9}$
& $2.3\times10^{-6}$
& $8.0\times10^{-11}$
& $3.3\times10^{-11}$
& $1.4\times10^{-10}$
& $2.0\times10^{-11}$
& $2.3\times10^{-11}$
& $2.8\times10^{-10}$
\\
\bottomrule
\end{tabular}%
}
\end{table*}

\begin{table*}[]
\centering
\caption{Peak-to-trough variations of the resonant fluxes for the orbits considered in Ref.~\cite{FHR}, with primary BH spin $a=0.90$, eccentricity $e=0.70$ and inclination $x_I=0.34202$, corresponding to $\theta_{\min}=20^\circ$ in the convention of Ref.~\cite{FHR}. For the energy, axial angular momentum and Carter constant fluxes, we report $\Delta_i^{\sigma}$ as defined in Eq.~\eqref{eq:DeltaFHR}, where $i\in\{E,L_z,Q\}$ labels the orbital integral and $\sigma\in\{\infty,H,\mathrm{tot}\}$ labels the flux channel.}
\label{tab:fhr_comparison_3}
\renewcommand{\arraystretch}{1.12}
\setlength{\tabcolsep}{3.2pt}
\resizebox{\textwidth}{!}{%
\begin{tabular}{
c c c c
c c c
c c c
c c c
}
\toprule
&
&
&
&
\multicolumn{3}{c}{Horizon $[\%]$}
& \multicolumn{3}{c}{Infinity $[\%]$}
& \multicolumn{3}{c}{Total $[\%]$}
\\
\cmidrule(lr){5-7}
\cmidrule(lr){8-10}
\cmidrule(lr){11-13}
Res.
& $e$
& $x_I$
& $p/M$
& $\Delta_{E}^H$
& $\Delta_{L_z}^H$
& $\Delta_{Q}^H$
& $\Delta_{E}^{\infty}$
& $\Delta_{L_z}^{\infty}$
& $\Delta_{Q}^{\infty}$
& $\Delta_{E}^{\rm tot}$
& $\Delta_{L_z}^{\rm tot}$
& $\Delta_{Q}^{\rm tot}$
\\
\midrule
$3{:}1$
& 0.7
& 0.34202
& 5.38952
& $4.0\times10^{0}$
& $3.8\times10^{-1}$
& $2.6\times10^{-1}$
& $6.1\times10^{-2}$
& $5.2\times10^{-2}$
& $5.8\times10^{-2}$
& $7.3\times10^{-2}$
& $1.1\times10^{-3}$
& $6.8\times10^{-2}$
\\
$2{:}1$
& 0.7
& 0.34202
& 6.31541
& $1.7\times10^{1}$
& $2.1\times10^{0}$
& $1.5\times10^{0}$
& $2.6\times10^{-1}$
& $1.9\times10^{-1}$
& $1.7\times10^{-1}$
& $2.8\times10^{-1}$
& $2.1\times10^{-2}$
& $2.0\times10^{-1}$
\\
$3{:}2$
& 0.7
& 0.34202
& 8.77436
& $9.7\times10^{1}$
& $2.0\times10^{1}$
& $1.1\times10^{1}$
& $1.3\times10^{0}$
& $7.0\times10^{-1}$
& $4.3\times10^{-1}$
& $1.2\times10^{0}$
& $1.2\times10^{-1}$
& $4.8\times10^{-1}$
\\
$4{:}3$
& 0.7
& 0.34202
& 11.42186
& $2.0\times10^{-4}$
& $1.4\times10^{-5}$
& $9.7\times10^{-5}$
& $1.7\times10^{-7}$
& $5.2\times10^{-8}$
& $4.0\times10^{-8}$
& $1.8\times10^{-8}$
& $2.3\times10^{-7}$
& $1.1\times10^{-7}$
\\
\bottomrule
\end{tabular}%
}
\end{table*}

\begin{table*}[]
\centering
\caption{Peak-to-trough variations of the resonant fluxes for the orbits considered in Ref.~\cite{FHR}, with primary BH spin $a=0.90$, eccentricity $e=0.70$ and inclination $x_I=0.93969$, corresponding to $\theta_{\min}=70^\circ$ in the convention of Ref.~\cite{FHR}. For the energy, axial angular momentum and Carter constant fluxes, we report $\Delta_i^{\sigma}$ as defined in Eq.~\eqref{eq:DeltaFHR}, where $i\in\{E,L_z,Q\}$ labels the orbital integral and $\sigma\in\{\infty,H,\mathrm{tot}\}$ labels the flux channel.}
\label{tab:fhr_comparison_4}
\renewcommand{\arraystretch}{1.12}
\setlength{\tabcolsep}{3.2pt}
\resizebox{\textwidth}{!}{%
\begin{tabular}{
c c c c
c c c
c c c
c c c
}
\toprule
&
&
&
&
\multicolumn{3}{c}{Horizon $[\%]$}
& \multicolumn{3}{c}{Infinity $[\%]$}
& \multicolumn{3}{c}{Total $[\%]$}
\\
\cmidrule(lr){5-7}
\cmidrule(lr){8-10}
\cmidrule(lr){11-13}
Res.
& $e$
& $x_I$
& $p/M$
& $\Delta_{E}^H$
& $\Delta_{L_z}^H$
& $\Delta_{Q}^H$
& $\Delta_{E}^{\infty}$
& $\Delta_{L_z}^{\infty}$
& $\Delta_{Q}^{\infty}$
& $\Delta_{E}^{\rm tot}$
& $\Delta_{L_z}^{\rm tot}$
& $\Delta_{Q}^{\rm tot}$
\\
\midrule
$3{:}1$
& 0.7
& 0.93969
& 3.27580 
& $1.3\times10^{-1}$
& $2.9\times10^{-1}$
& $5.2\times10^{-1}$
& $3.0\times10^{-2}$
& $2.5\times10^{-2}$
& $5.3\times10^{-2}$
& $2.7\times10^{-2}$
& $1.4\times10^{-2}$
& $3.2\times10^{-2}$
\\
$2{:}1$
& 0.7
& 0.93969
& 3.78947
& $4.7\times10^{-1}$
& $9.2\times10^{-1}$
& $2.8\times10^{0}$
& $9.4\times10^{-2}$
& $6.6\times10^{-2}$
& $1.6\times10^{-1}$
& $8.4\times10^{-2}$
& $3.8\times10^{-2}$
& $6.7\times10^{-2}$
\\
$3{:}2$
& 0.7
& 0.93969
& 5.48622
& $6.2\times10^{0}$
& $4.7\times10^{0}$
& $2.2\times10^{1}$
& $3.1\times10^{-1}$
& $1.4\times10^{-1}$
& $3.5\times10^{-1}$
& $2.3\times10^{-1}$
& $6.8\times10^{-2}$
& $3.9\times10^{-3}$
\\
$4{:}3$
& 0.7
& 0.93969
& 7.53814
& $3.8\times10^{-7}$
& $6.7\times10^{-7}$
& $1.7\times10^{-5}$
& $5.8\times10^{-9}$
& $2.6\times10^{-9}$
& $3.4\times10^{-8}$
& $3.9\times10^{-9}$
& $4.0\times10^{-9}$
& $3.1\times10^{-8}$
\\
\bottomrule
\end{tabular}%
}
\end{table*}

\bibliographystyle{apsrev4-1}
\bibliography{resonance_coeffs}

\end{document}